\documentclass[10pt,aps,twocolumn,floatfix,showpacs,showkeys,preprintnumbers,nofootinbib,superscriptaddress,longbibliography,bibnotes]{revtex4-1}
\pdfoutput=1
\usepackage[colorlinks=true,breaklinks=true]{hyperref}
\usepackage[utf8]{inputenc}
\hypersetup{allcolors=[rgb]{0.0 0.0 0.6},linkcolor=[rgb]{0.75 0.05 0.05}}
\usepackage{amsmath,amssymb}
\usepackage{epsfig}  
\usepackage{graphicx}   
\usepackage{slashed}             
\usepackage{url}
\usepackage{color}
\usepackage{multirow}
\usepackage{placeins}
\usepackage[dvipsnames]{xcolor}
\usepackage{letltxmacro}
\LetLtxMacro{\oldcite}{\cite}
\renewcommand{\cite}[1]{\mbox{\oldcite{#1}}}

\allowdisplaybreaks

\bibpunct{[}{]}{,}{n}{}{,}

\begin{document}

\title{Fast neutrino flavor instability in the neutron-star convection layer of three-dimensional
supernova models}

\author{Robert~Glas}
\email{rglas@mpa-garching.mpg.de}
\affiliation{Max-Planck-Institut f\"ur Astrophysik, Karl-Schwarzschild-Stra{\ss}e 1, 85748 Garching, Germany}
\affiliation{Physik Department, Technische Universit\"at M\"unchen, James-Franck-Stra{\ss}e 1, 85748 Garching, Germany}
\author{H.-Thomas~Janka}
\email{thj@mpa-garching.mpg.de}
\affiliation{Max-Planck-Institut f\"ur Astrophysik, Karl-Schwarzschild-Stra{\ss}e 1, 85748 Garching, Germany}
\author{Francesco Capozzi}
\email{capozzi@mppmu.mpg.de}
\affiliation{Max-Planck-Institut f\"{u}r Physik (Werner-Heisenberg-Institut), F\"{o}hringer Ring 6, 80805 M\"{u}nchen, Germany 
}                       
\author{Manibrata~Sen}
\email{manibrata@berkeley.edu }
\affiliation{Department of Physics, University of California Berkeley, Berkeley, California 94720, USA}
\affiliation{Department of Physics and Astronomy, Northwestern University, Evanston, IL 60208, USA}
\author{Basudeb~Dasgupta}
\email{bdasgupta@theory.tifr.res.in}
\affiliation{Tata Institute of Fundamental Research,
             Homi Bhabha Road, Mumbai, 400005, India}
\author{Alessandro~Mirizzi}
\email{alessandro.mirizzi@ba.infn.it }
\affiliation{Dipartimento Interateneo di Fisica ``Michelangelo Merlin'', Via Amendola 173, 70126 Bari, Italy}
\affiliation{Istituto Nazionale di Fisica Nucleare - Sezione di Bari,
Via Amendola 173, 70126 Bari, Italy}
\author{G\"unter~Sigl}
\email{guenter.sigl@desy.de}  
\affiliation{II. Institute for Theoretical Physics, Hamburg University,
Luruper Chaussee 149, 22761 Hamburg, Germany}


\begin{abstract}
It has been speculated for a long time that neutrinos from a supernova (SN) may undergo fast flavor 
conversions near the collapsed stellar core. We perform a detailed study of this intriguing possibility, 
for the first time analyzing two time-dependent
state-of-the-art three-dimensional (3D) SN models of 9\,$M_\odot$ and 20\,$M_\odot$
from recent papers of Glas et al.
Both models were computed with multi-dimensional three-flavor neutrino transport based on a two-moment 
solver, and both exhibit the presence of the so-called lepton-number emission self-sustained asymmetry
(LESA).
The transport solution does not provide the angular distributions of the flavor-dependent neutrino 
fluxes, which are crucial to track the fast flavor instability. To overcome this limitation,
we use a recently proposed approach based on the angular moments of the energy-integrated electron 
lepton-number distribution up to second order, i.e., angle-energy integrals of the difference
between $\nu_e$ and $\bar\nu_e$ phase-space distributions multiplied by corresponding powers of the 
unit vector of the neutrino velocity. With this method we find the possibility of fast
neutrino flavor instability at radii smaller than $\sim$20\,km, which is well interior to the 
neutrinosphere where neutrinos are still in the diffusive and near-equilibrium regime. Our results 
confirm recent observations in a 2D (axisymmetric) SN model and in 2D and 3D models with fixed
matter background, which were computed with Boltzmann neutrino transport. However, the flavor
unstable locations are not isolated points as discussed previously, but thin skins surrounding volumes
where $\bar\nu_e$ are more abundant than $\nu_e$. These volumes grow with time and appear first in the 
convective layer of the proto-neutron star (PNS), where a decreasing electron fraction and 
high temperatures favor the occurrence of regions with negative neutrino chemical potential.
Since the electron fraction remains higher in the LESA dipole direction, where convective lepton-number
transport out from the nonconvective PNS core slows down the deleptonization, flavor unstable
conditions become more widespread in the opposite hemisphere. This interesting phenomenon 
deserves further investigation, since its impact on SN modeling and possible consequences for 
SN dynamics and neutrino observations are presently unclear.
\end{abstract}

\maketitle

\section{Introduction}

The deepest supernova (SN) regions provide a unique laboratory to probe neutrino flavor conversions 
in a non-linear regime, where the neutrino evolution is determined mainly by their mutual interactions. 
Indeed, at distances $r \lesssim {\mathcal O}(10^2)$\,km from the centre of the SN, the neutrino 
density $n_{\nu}$ is so high that it dominates the flavor evolution, leading to self-induced neutrino flavor
 conversions~\cite{Pantaleone:1992eq,Kostelecky:1994dt,Pastor:2002we}. 
These  have been a topic of intense investigation for over a 
decade~\cite{Sawyer:2005jk, Duan:2006an, Hannestad:2006nj, Fogli:2007bk, Sawyer:2008zs, Dasgupta:2009mg, Sawyer:2015dsa}. 
See refs.\,\cite{Duan:2010bg,Mirizzi:2015eza,Chakraborty:2016yeg,Horiuchi:2017sku} for recent reviews.

In this context, a peculiar type of  self-induced flavor conversions, called ``fast'' instabilities~\cite{Sawyer:2005jk,Sawyer:2008zs,Sawyer:2015dsa,Chakraborty:2016lct,Dasgupta:2016dbv,Izaguirre:2016gsx,Capozzi:2017gqd, Dighe:2017sur, Dasgupta:2017oko,Abbar:2017pkh,Airen:2018nvp,Abbar:2018beu,Yi:2019hrp,Capozzi:2019lso,ShalgarTamborra2019,Martin:2019gxb,Martin:2019dof}, is expected to lead to flavor conversions developing on very short distances,  $r \lesssim {\mathcal O}(1)$\,m. 
Fast flavor conversions have been associated with ``crossings" in the electron lepton number (ELN) angular distribution, i.e., with a change of sign in the difference between $\nu_e$ and $\bar{\nu}_e$ number densities as a function of emission
 angle~\cite{Dasgupta:2016dbv,Capozzi:2017gqd,Yi:2019hrp}.
Conditions for crossings in the ELN were expected to be possible in the neutrino decoupling 
region in SN cores, where the different flavors have significantly different angular distributions.

This possibility of fast flavor conversions and its potential effects on SN dynamics and nucleosynthesis has stimulated several studies to assess the occurrence  of fast instabilities in different SN models. A first study 
 in this direction was performed  in~\cite{Tamborra:2017ubu}, where a dedicated analysis of the angular distributions of the neutrino radiation field  for several spherically symmetric (1D) SN simulations has not found any crossing  in the ELN near the neutrinosphere. 
More generally, 2D or 3D models can exhibit a large-scale dipole in the ELN emission, termed Lepton-Emission Self-sustained Asymmetry (LESA)~\cite{Tamborra:2014aua}, which also makes a crossing more likely to occur. 
In this context, the first analyses of fast instabilities in multidimensional SN models have been recently 
 performed in~\cite{Abbar:2018shq,Azari:2019jvr}. 
 In~\cite{Abbar:2018shq} the authors   extracted three snapshots from numerical data  in 2D and 3D SN simulations and looked for ELN crossings in the angular distributions of $\nu_e$ and $\bar\nu_e$. They found favorable conditions in extended regions with the radius of   $50$--$70$~km. 
 Then, by a linear stability analysis of the neutrino equations of motion, they identified the strength of this instability for a representative point.
However, their neutrino distributions were obtained from neutrino transport calculations done in a post-processing step, dropping the time dependence of both hydrodynamical and neutrino quantities and ignoring matter motions entirely. Hence they are not fully self-consistent.
 Conversely, in~\cite{Azari:2019jvr}  the authors used fully self-consistent simulations in 2D, computing neutrino transport with a multi-angle Boltzmann solver coupled to hydrodynamics.
 Applying linear stability analysis near the neutrinosphere they found no positive signatures of conversion at least for the spatial points and times studied in ther particular model. 

Furthermore, in~\cite{Morinaga+2019} the claim was
made that in the pre-shock region, at $r \simeq {\mathcal O}(100)$\,km, the residual coherent 
neutrino-nucleus scatterings could produce a tiny crossing in the ELN, whose presence has been
confirmed by the inspection of various numerical simulations.
Despite the smallness of the crossing, according to a stability analysis  
it would be enough to trigger significant fast conversions. However, for a
cautioning argument against overinterpreting results of stability analyses, 
see~\cite{Shalgar+2019}.

Recently, two publications, based again on the SN models considered in \cite{Abbar:2018shq,Azari:2019jvr},
reported positive detections of 
locations of ELN crossings deep inside the proto-neutron star (PNS) when investigating the selfconsistent 
2D core-collapse simulation of an 11.2\,M$_\odot$ star computed with Boltzmann neutrino transport 
\cite{DelfanAzari+2019} and 3D Boltzmann neutrino-transport results for fixed matter background at some
instants during the post-bounce evolution of 11.2\,M$_\odot$ and 27\,M$_\odot$ progenitors
\cite{Abbar+2019}. Because the diffusive conditions for neutrinos in the deep PNS interior imply that
the angular distributions of both $\nu_e$ and $\bar\nu_e$ are nearly isotropic, ELN crossings
were found only in regions where the ``asymmetry parameter'' $\Gamma = n_{\bar\nu_e}/n_{\nu_e}$,
i.e.\ the ratio of the number densities of both neutrino types, is close to unity (see also
\cite{ShalgarTamborra2019}). Consequently
and naturally in the equilibrium diffusion regime, the chemical potential of $\nu_e$ 
nearly vanishes in these regions. The authors of \cite{DelfanAzari+2019} speculated that the appearance 
of light nuclei (among them $\alpha$ particles as the dominant species) is causal for the
development of such instability conditions.
Just as Delfan Azari et al.\ \cite{DelfanAzari+2019}, also
Abbar et al.\ \cite{Abbar+2019} diagnosed ELN crossings in deep regions inside the PNS 
only in a small number of isolated points at the analysed post-bounce moments. 
They also correlated their occurrence with locations where the chemical potential of 
electron neutrinos nearly vanishes and pointed out that the electron fraction $Y_e$ 
is relatively low there and the temperature is close to maximal values.

These interesting results motivate the need to extend the search for fast instabilities to other 
state-of-the-art and fully self-consistent multidimensional SN models.
However, most multi-D SN simulations~\cite{Tamborra:2014aua,Bruenn:2014qea, OConnor:2015rwy, Nagakura:2017mnp, Richers:2017awc, Vartanyan:2018xcd, Just:2018djz, Cabezon:2018lpr, Pan:2018vkx,Glas:2018vcs,Glas:2018oyz} evolve only the (energy-dependent) angular integrals (``moments'') of the neutrino phase-space distributions with time, and not the fully angle-dependent distributions. Reference~\cite{Nagakura:2017mnp}, used in~\cite{Azari:2019jvr,DelfanAzari+2019}, is a welcome exception. The lack of detailed angular information seems to preclude a linear stability analysis that requires knowing these distributions.  To overcome this limitation, some of us have recently proposed an alternative method  to diagnose the possibility of fast instabilities in the absence of detailed knowledge of the ELN distributions~\cite{Dasgupta:2018ulw}. This recipe is based on identifying a specific Fourier mode of the flavor instability field called the ``zero mode'', which has an easily calculable growth rate depending only on the angular moments of the ELN up to second order. It has been shown with numerical examples that the growth rate of this mode, calculated from the stability analysis, nicely approximates the growth of flavor conversions for the same mode in detailed numerical calculations.
 
The purpose of the present work is to use this new method of analysis to scan the different regions in self-consistent, state-of-the-art 3D SN models with fully 3D two-moment neutrino transport for the possibility of fast flavor conversions therein. Specifically, we employ two time-dependent stellar core-collapse (and explosion) simulations
for 9\,M$_\odot$ and 20\,M$_\odot$ progenitors recently published 
by the Garching group~\cite{Glas:2018vcs,Glas:2018oyz}. Both of these simulations exhibit the LESA phenomenon. We will demonstrate that the direct evaluation of discretized numerical data provided by computational models leads to the identification of only few, isolated points of ELN crossings. We will argue that instead of being such point-like locations, the regions of fast flavor instability are thin 2D layers that first appear around small 3D volumes in the convective layer of the PNS, and which grow with time as the convective and diffusive transport of electron-lepton number drives a decrease of $Y_e$ in the PNS convection layer. We also observe a strong hemispheric asymmetry of the thin layer of flavor instability correlated with the asymmetry of PNS convection leading to the LESA phenomenon. The regions of ELN crossings are much more extended in the hemisphere opposite to the LESA dipole direction, where PNS convection is weaker and $Y_e$ is lower. Our analysis thus shows that the locations of ELN crossings are not dot-like and fluctuating because of stochastic hydrodynamical variations, but they are long-lasting and large-scale structures (for possibly important implications of this fact, see \cite{Shalgar+2019}).

In Sec.~\ref{sec:2}, we describe our method for diagnosing instabilities based on the angular
moments of the neutrino phase-space distributions.
In Sec.~\ref{sec:3} we present the results of our search for fast flavor instability 
in the two investigated 3D SN models, first by directly using the discretized output
of the numerical simulations in the instability condition 
(Sec.~\ref{sec:3A}), which leads to the identification of
only few isolated points of flavor instability located interior to the neutrinospheres 
in the convective shell of the PNS. In Sec.~\ref{sec:3B} we discuss the conditions
for ELN crossings. We argue that this direct analysis on the discrete numerical mesh
fails to correctly identify the regions of flavor instability, which are actually thin 2D
layers in 3D space. We propose an alternative, better strategy to find these layers
containing the locations of flavor instability. In Sec.~\ref{sec:4} we present the time
evolution of the instability layers in our two model runs (Sec.~\ref{sec:4A})
and explain the reason for the development of the relevant physical conditions
(Sec.~\ref{sec:4B}). We also discuss
their correlation with the dipole of the lepton-number emission and the asymmetry of the 
electron distribution in the PNS convection layer connected with the LESA phenomenon.
Finally, in Sec.~\ref{sec:5}, we conclude with a brief summary.

\begin{figure}[!t]
\begin{centering}
\includegraphics[width=\columnwidth]{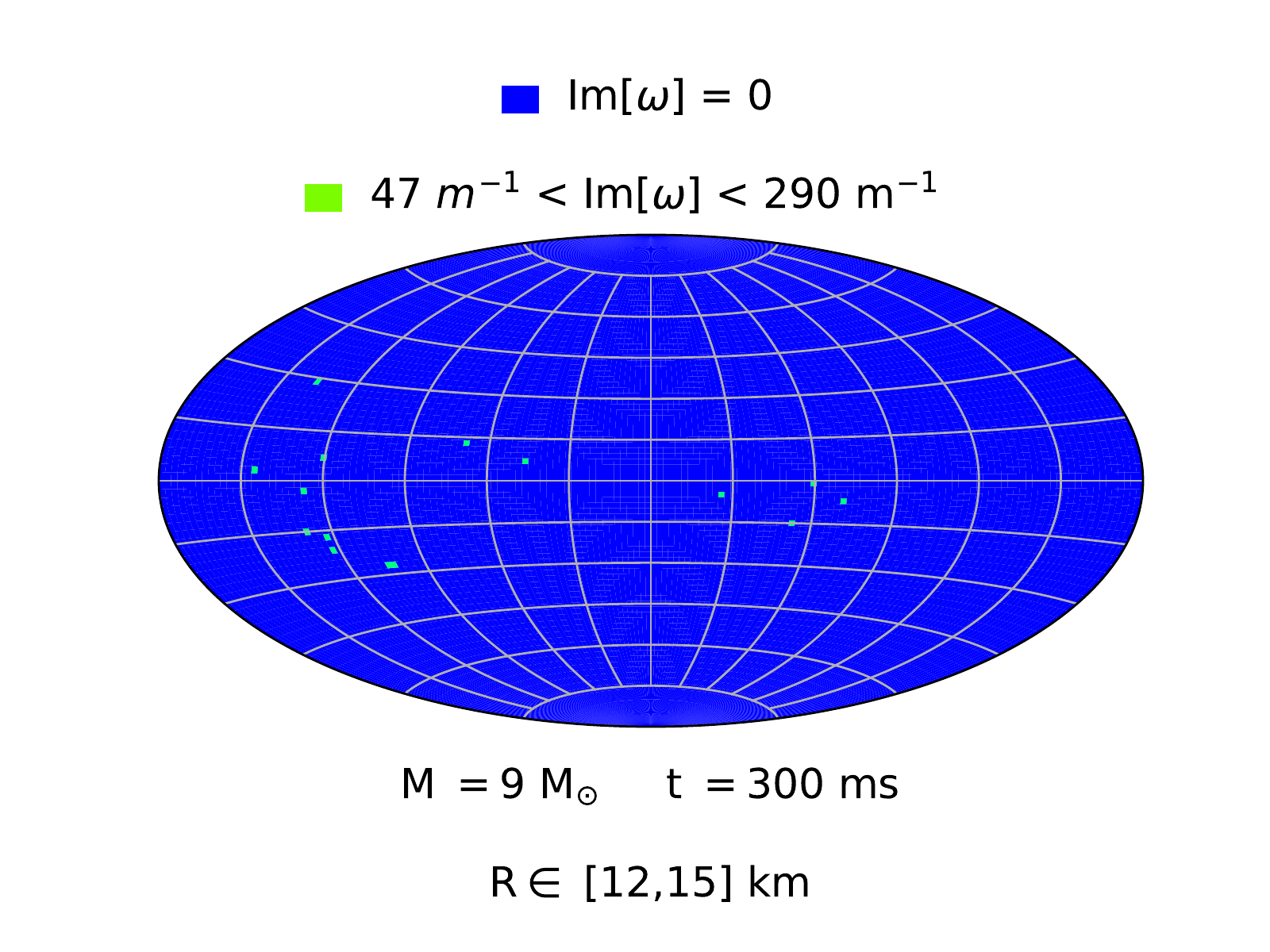}\\
\includegraphics[width=\columnwidth]{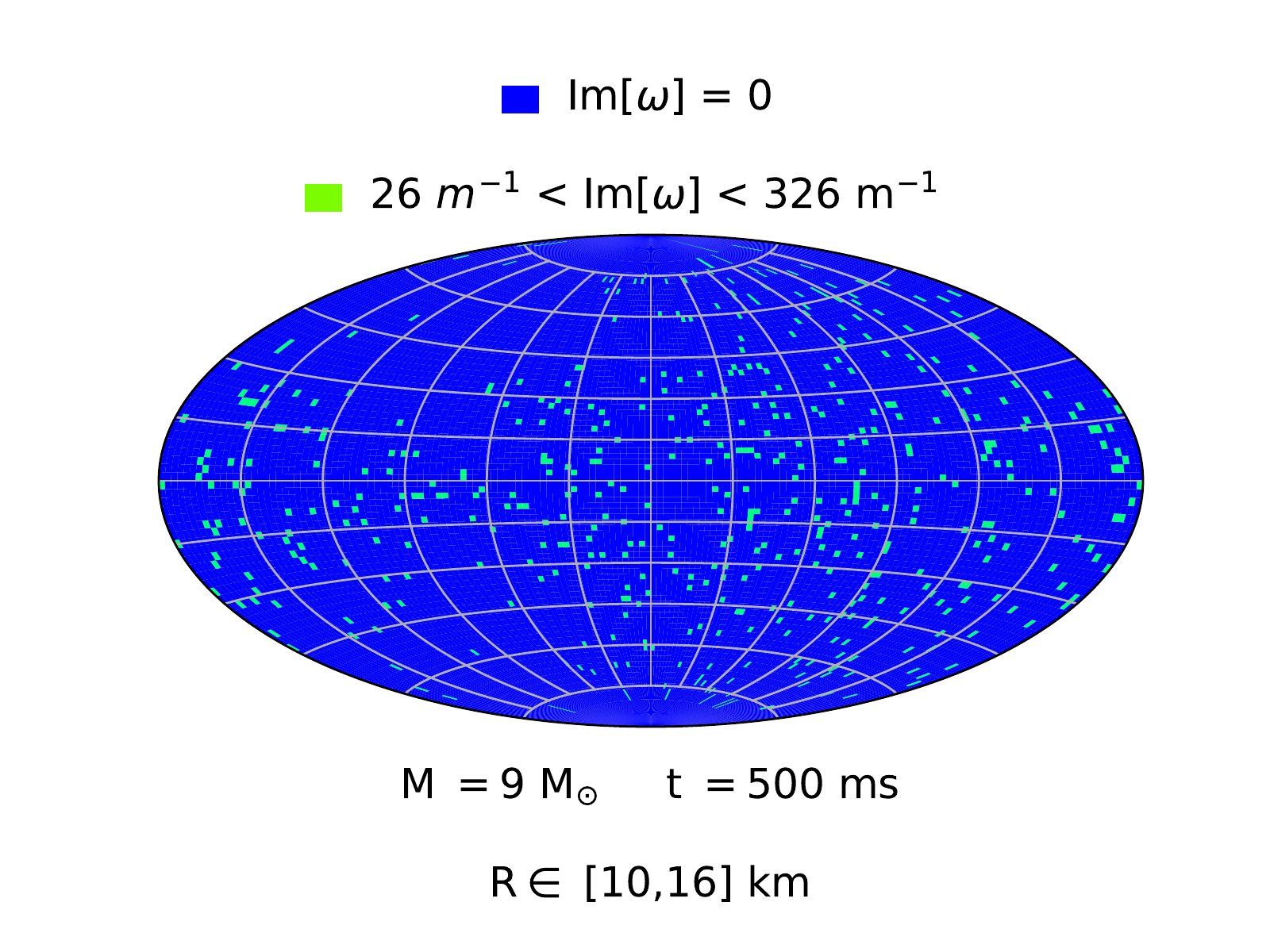}\\
\includegraphics[width=\columnwidth]{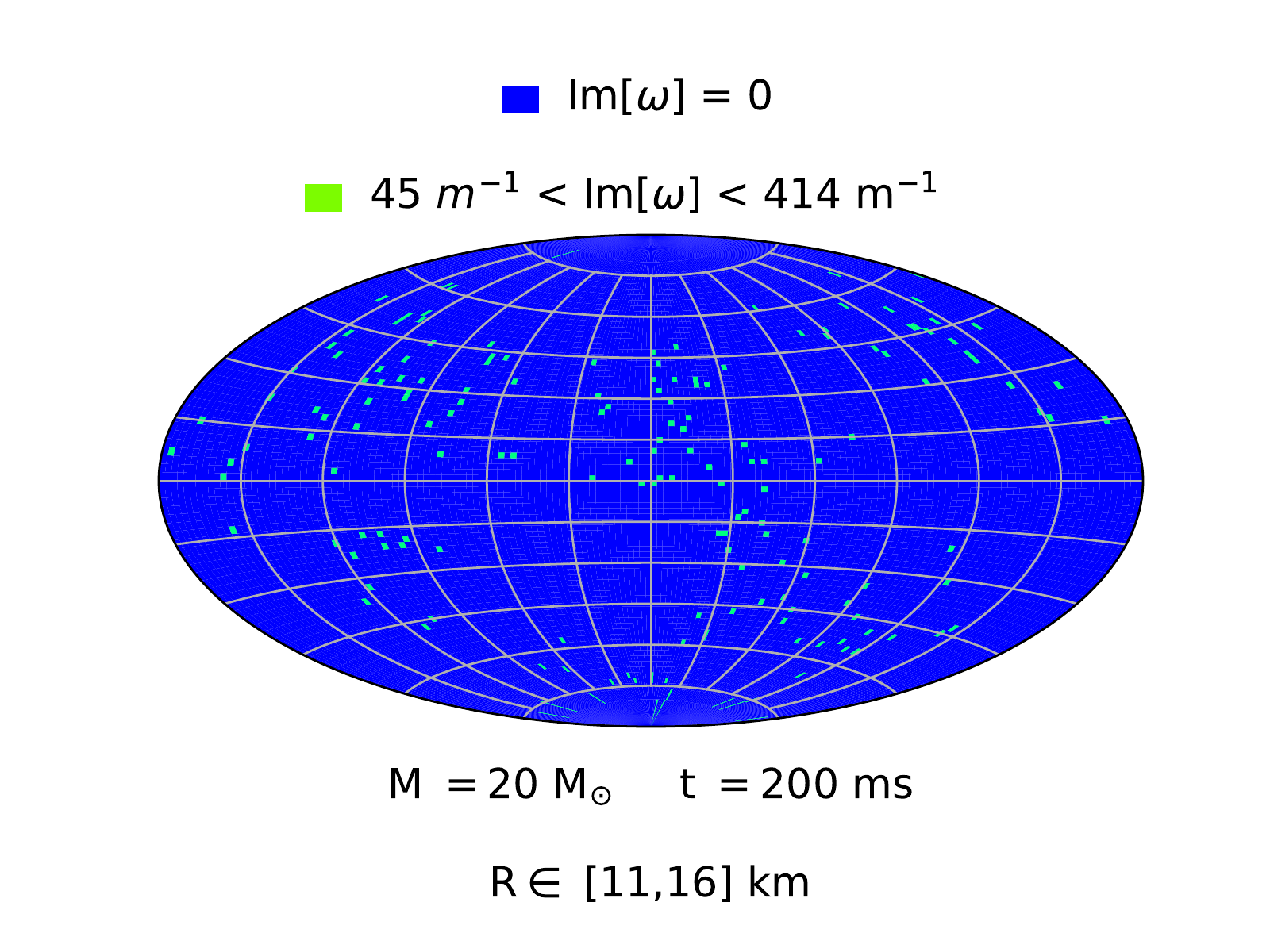}
\end{centering}
\caption{Aitoff projections for the $\log_{10}{\rm Im}[\omega]$ obtained by solving
Eq.~(\ref{eq:disp0}) in the case of the 9\,$M_{\odot}$ star
at $t=300$\,ms and 500\,ms after core bounce ({\em upper two panels}) and of the 20\,$M_{\odot}$ star
at 200\,ms post bounce ({\em bottom panel}). The plots are
obtained by selecting the largest $\log_{10}{\rm Im}[\omega]$ in the radial range of
$[10,30]$\,km.}
\label{fig:Im_omega_numerical}
\end{figure}

\section{fast  instabilities and  moments of neutrino distributions}
\label{sec:2}

\subsection{Instability equation}

In order to track the existence of the fast neutrino flavor instability one has to 
perform a linear stability analysis of the neutrino equations of motion. 
We refer the interested reader to \cite{Airen:2018nvp} for a detailed discussion 
of this analysis and for the related publications.
Here, we simply mention that in the growing literature about fast flavor conversions, a novel 
approach to study these effects was recently proposed in\,\cite{Izaguirre:2016gsx}. This  is based on the \emph{dispersion relation} for the frequency and wavenumber 
$(\omega, {\bf k})$ in the mean field of $\nu_e\nu_x$ coherence, which is 
essentially the off-diagonal element of the neutrino density matrix $\varrho({\bf p}, {\bf x},t)$ that we will call $S$ in the following.
One looks for solutions of
the linearized equations for the flavor evolution in the
form 
\begin{equation}
S \sim  e^{i({\bf k} \cdot {\bf x} - \omega t)} \,\ .
\label{eq:wave0}
\end{equation}
Typically such a solution may exist only if $\omega$ and ${\bf k}$ are related by an appropriate equation, called the \emph{dispersion relation}.
Loosely speaking, if either $ {\bf k}$ or $\omega$ develop imaginary parts leading to positive real arguments in the exponential, the solution  
is expected  to grow in space or time, thus signaling  an ``instability''. Unfortunately, identifying the instabilities rests upon a more complicated analysis that requires a full characterization of the complex analytic structure of the dispersion relation~\cite{Capozzi:2017gqd}.

In~\cite{Dasgupta:2018ulw} some of us proposed a simpler analytical tool to diagnose the fast neutrino instability.
Our proposal is based on identifying a specific Fourier mode of the flavor instability field that we call the ``zero mode''
in a corotating frame, in which one can gauge-away the matter term from the neutrino equations of motion.
We labeled this zero-mode as ${\bf k}=0$.~\footnote{A remark is in order. Our method  does not completely exclude the presence of an instability for those points where we 
find  ${\rm Im}(\omega)=0$. Indeed, as our analysis is based only on the zero mode ${\bf k}=0$, 
there might be  a ${\bf k}\ne0$ that is unstable. Therefore, 
one cannot exclude larger instability regions than those that we will show. 
The reader is referred to \cite{Johns+2019} for a comparison of instability 
regions obtained with different instability criteria.}

This is motivated by the fact that the calculation of $\omega$ for this mode is significantly simpler than a full characterization of the dispersion relations, $D(\omega, {\bf k})$~\cite{Capozzi:2017gqd}. In fact, for this mode the
dispersion relation becomes
\begin{equation}
D(\omega,0)=\textrm{det}\left(\eta^{\mu\nu}+\frac{1}{\omega}\,V^{\mu\nu}\right)=0
\label{eq:disp0}
\end{equation}
with $\eta^{\mu\nu} = \mathrm{diag}(+1,-1,-1,-1)$,
i.e., $D(\omega,0)$ is a polynomial in $\omega$. The specific model of SN neutrino populations and their angular distributions, encoded in the ELN, only enters the equation through the tensor  $V^{\mu\nu}$  (with $\mu, \nu=0,1,2,3$) that contains the angular moments of the neutrino distributions up to second order in the neutrino velocity, namely,
\begin{equation}
V^{\mu\nu} = \int \frac {d {\bf v}}{4 \pi}\, {v^\mu v^\nu} G_{\bf v}\,\ , \\
\label{eq:momenta}
\end{equation}
where $v^\mu=(1,{\bf p}/E)$, i.e.  the zeroth component of the velocity four-vector is 1 and the spatial components are given by the unit vector ${\bf v}={\bf p}/E$
(with $E=|{\bf p}|$).
The function
\begin{equation}
G_{\bf v} = \sqrt{2} G_F \int_{0}^{\infty}\frac{dE\,E^2}{2 \pi^2}\left[f_{\nu_e}(E,{\bf v})-f_{\bar\nu_e}(E,{\bf v}) \right] 
\label{eq:eln}
\end{equation}
is the difference of the phase-space occupation functions integrated over energy space, 
i.e., the angular distribution of the electron-neutrino lepton number (ELN)~\cite{Chakraborty:2016lct}.
Here we assume that $\nu_x$ and $\bar{\nu}_x$ have identical distributions.
Therefore, $V^{\mu\nu}$
depends on the angular moments of the neutrino-species dependent phase space distributions up to
second order in the neutrino velocities, i.e.,
\begin{equation}
V^{\mu\nu}= \langle {v^\mu v^\nu} \rangle_{\nu_e} - \langle {v^\mu v^\nu} \rangle_{\bar{\nu}_e}\,, 
\end{equation}
where the notation $\langle \ldots \rangle_{\nu_\alpha}$ refers to
\begin{equation}
\langle \ldots \rangle_{\nu_\alpha} \equiv \sqrt{2} G_F \int \frac{d^3{\bf p}}{(2\pi)^3}\, (\ldots)\,f_{\nu_\alpha}({\bf p})\,.
\label{eq:momflav}
\end{equation}
These (energy-integrated) moments are related to the difference of
the number densities ($n_\nu$) of $\nu_e$ and $\bar\nu_e$, the
corresponding difference of the number-flux densities ($F_\nu^r$), and the 
difference of the second angular moments of the $\nu_e$ and $\bar\nu_e$ 
number distributions ($P_\nu^{rr}$) as follows:
\begin{eqnarray}
V^{00}\,(\sqrt{2}G_F)^{-1} &=&n_{\nu_e}-n_{\bar{\nu}_e} \equiv \Delta n_\nu \,,\label{eq:equivalence_moments1}\\
V^{0r}\,(\sqrt{2}G_F)^{-1} &=&(F_{\nu_e}^r-F_{\bar{\nu}_e}^r)\,c^{-1} \equiv \Delta F_\nu\,c^{-1} \,,\label{eq:equivalence_moments2}\\
V^{rr}\,(\sqrt{2}G_F)^{-1} &=&P_{\nu_e}^{rr}-P_{\bar{\nu}_e}^{rr} \equiv \Delta P_\nu^{rr} \,.\label{eq:equivalence_moments3}  
\end{eqnarray}
where for later quantitative evaluation we have reintroduced
the factors $c$ in the expressions in the rhs of these relations.

We remind the reader that the two-moment ``M1'' neutrino-transport scheme used in 
\cite{Glas:2018vcs,Glas:2018oyz} evolves
the ``$00$'' and the  ``$0i$'' components (with $i \in \{r, \theta, \phi \}$ being
radius $r$, polar angle $\theta$, and  azimuthal angle $\phi$ of the polar coordinate
system) of the moments in Eq.~(\ref{eq:momflav}) in time for all neutrino species.
The system of moment equations of the transport solver is closed by an algebraic 
relation for non-evolved moments (i.e., for the ``$ij$'' components of the tensor
with $i\ge 1$ and $j\ge 1$),
which depend on the ``$00$'' and the ``$0i$'' components (see e.g. \cite{Murchikova:2017zsy}).


\begin{figure*}[!t]
\begin{centering}
\includegraphics[width=2\columnwidth]{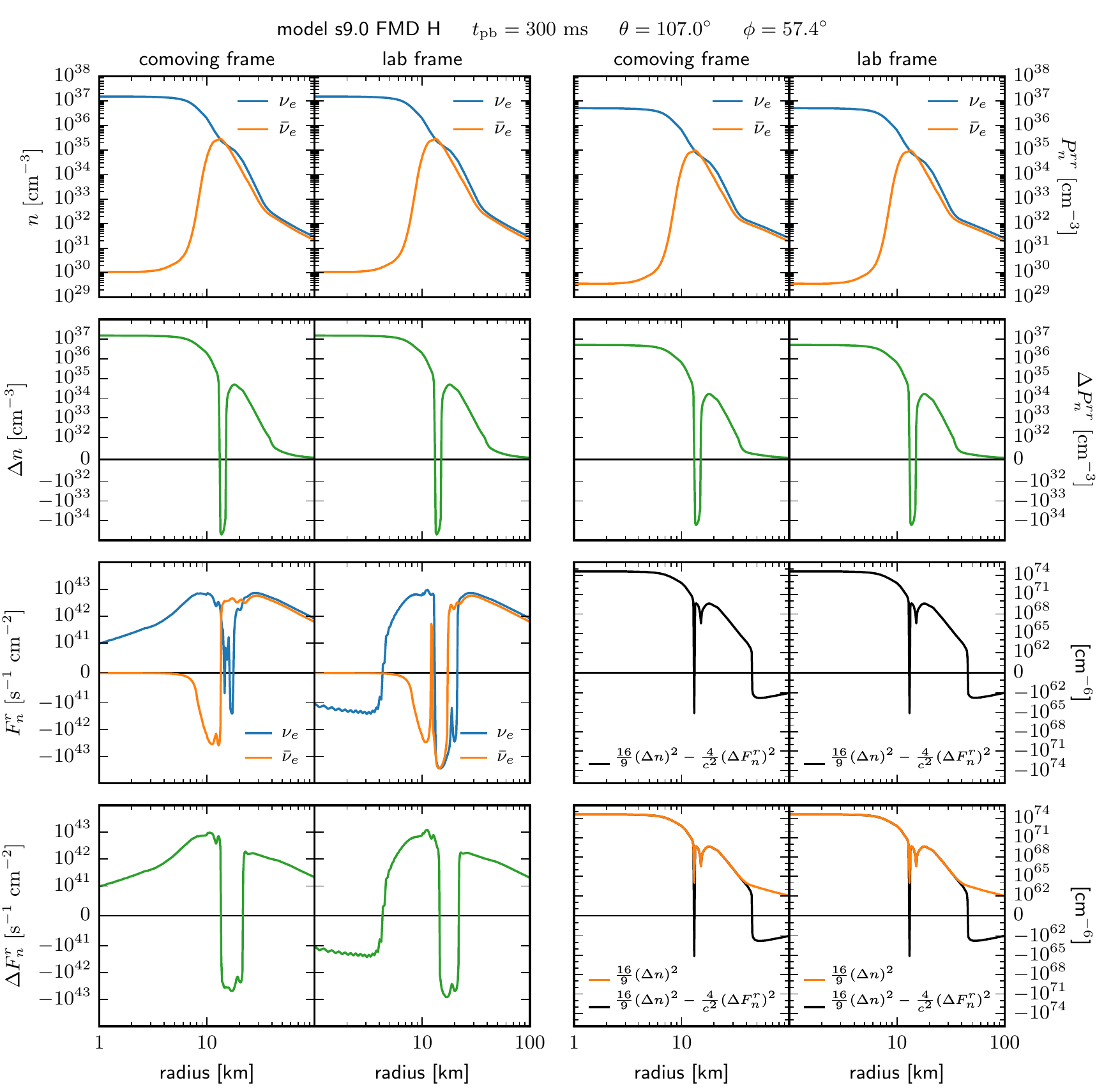}
\end{centering}
\caption{Radial profiles of the basic (energy-integrated) angular moments
$n_\nu$, $F_\nu^r$, and $P_\nu^{rr}$ for $\nu_e$ and $\bar\nu_e$ as well
as their differences $\Delta n_\nu$, $\Delta F_\nu^r$, and $\Delta P_\nu^{rr}$ 
along a radial direction at $(\theta,\phi) = (107^\circ,57.4^\circ)$ in our
9\,M$_\odot$ model at 300\,ms after bounce. The {\em four panels on the lower right}
display the corresponding flavor-instability functional ${\cal F}$ and, for
comparison, the first term of it, $\frac{16}{9}\,(\Delta n_\nu)^2$, as
labeled in the panels.}
\label{fig:moments-radius}
\end{figure*}
\begin{figure}[t!]
\begin{centering}
\includegraphics[width=\columnwidth]{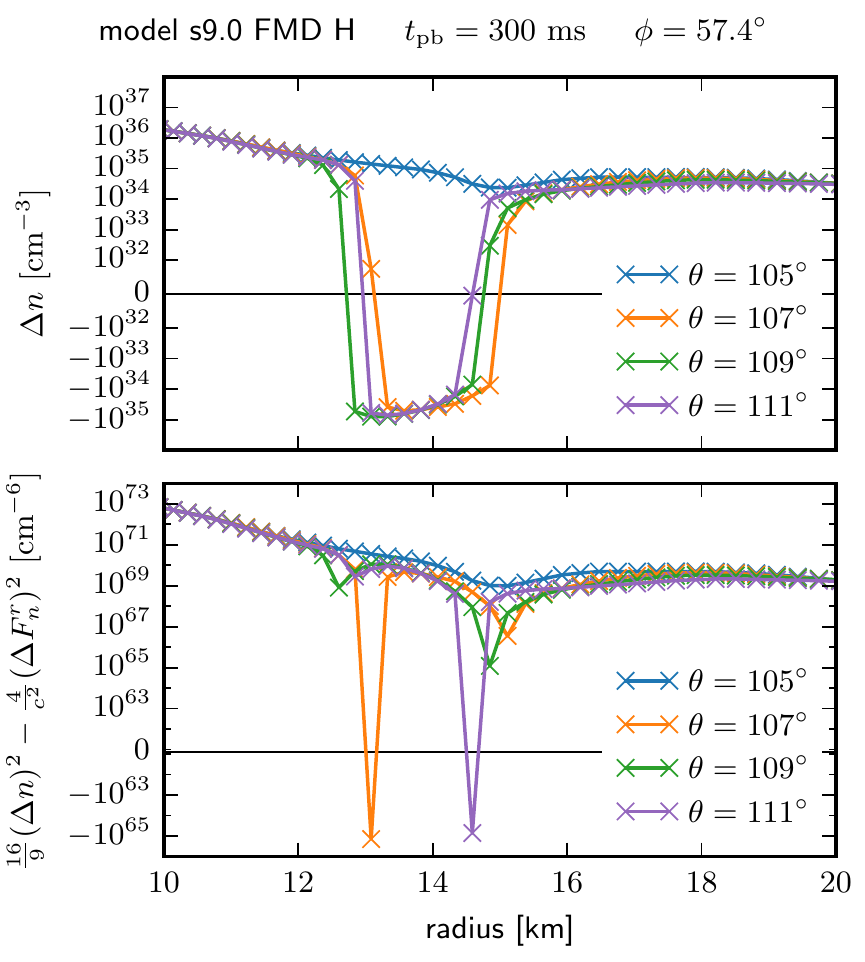}
\end{centering}
\caption{Radial profiles of $\Delta n_\nu$ ({\em upper panel}) 
and ${\cal F}$ ({\em lower panel}) in the vicinity of the two local
minima of ${\cal F}$ visible around $\sim$14\,km in
Fig.~\ref{fig:moments-radius} in the case of our 9\,M$_\odot$ simulation
at 300\,ms after bounce. Besides the radial direction chosen for 
Fig.~\ref{fig:moments-radius}, three other radial directions 
with neighboring zenith angles for fixed azimuthal angle of
$\phi=57.4^\circ$ are shown, too. The individual radial mesh
points of the computational grid are marked by crosses.}
\label{fig:Instability_criterion}
\end{figure}

\section{Search for  fast instabilities}
\label{sec:3}

\subsection{Direct analysis of discretized numerical results}
\label{sec:3A}

In this work we employ energy-integrated angular moments of the neutrino number distribution 
provided by the neutrino transport solver used in the considered 3D SN simulations of 
Refs.~\cite{Glas:2018oyz,Glas:2018vcs}, Models s9.0\,FMD\,H and s20\,FMD\,H there.
Appropriate normalization constants as specified in 
Eqs.~(\ref{eq:equivalence_moments1})--(\ref{eq:equivalence_moments3})
are applied in order to match the quantities in Eq.~(\ref{eq:momenta}). 
We choose to work mainly in the comoving frame of the stellar fluid (fluid frame), where SN neutrino transport usually provides its output quantities.
Equivalent results in terms of flavor instabilities are obtained in the laboratory 
frame, i.e. the rest frame of the stellar center. We will briefly demonstrate this later.

We mainly focus on our progenitor with 9\,M$_\odot$ and discuss similarities as
well as differences compared to the 20\,M$_\odot$ simulation. 
Figure~\ref{fig:Im_omega_numerical} displays Aitoff projections of the 9\,M$_\odot$ model
for post-bounce times of $t=300$\,ms and 500\,ms (upper two panels) and of the 
20\,M$_\odot$ star for $t=200$\,ms after bounce (bottom panel). Different radial directions 
in the 3D simulations correspond to discretized values of the zenith angle $\theta$ and
azimuthal angle $\phi$. Points ($\phi,\theta$) are marked by green color if the solutions
of Eq.~(\ref{eq:disp0}) yield
${\rm Im}(\omega) > 0$ for at least one value of the discretized radial coordinate $r$
in the range $10\,\mathrm{km}\le r \le 30\,\mathrm{km}$. 
If the color is blue then ${\rm Im}(\omega) = 0$ for all radii, i.e. there is no instability. 

In the 9\,M$_\odot$ model at 300\,ms
we find instability points only in the interval of $[12,15]$ km and for very specific directions, 
i.e., a few isolated and unconnected pairs of values ($\phi,\theta$). As specified in the plots, 
${\rm Im}(\omega)\sim O(10-100)$\,m$^{-1}$ at unstable locations, which means the flavor 
instability can develop over a time scale even shorter than a nanosecond.
In the middle and lower panels of Figure~\ref{fig:Im_omega_numerical} we witness a larger number 
of points of instability than in the upper panel. 

Nonzero values of Im($\omega$) in the 9\,M$_\odot$ model occur only later than $\sim$300\,ms
after bounce, whereas in the 20\,$M_\odot$ the condition for flavor instability shows up 
earlier and more points with positive imaginary part of $\omega$ are present already at 200\,ms.

We note in passing that we have not detected any points of flavor instability between the
neutrino-decoupling region and the SN shock (at $r \ge 30$--50\,km, depending on the
post-bounce time). Some authors have speculated about the possibility that ELN crossings 
might occur in this region in multi-dimensional models and, indeed, Ref.~\cite{Abbar:2018shq} 
as well as Refs.~\cite{Abbar+2019,Nagakura+2019} have reported such locations in 
2D and 3D simulations of an 11.2\,M$_\odot$ star. In this context, however, it is important
to keep in mind that our moment-based criterion employs differences of angle and energy 
integrals of the neutrino distribution functions. Such an integral criterion is not
necessarily sensitive enough to diagnose low-level crossings in the angular
distributions of $\nu_e$ and $\bar\nu_e$, i.e., small differences in the distribution
functions leading to a reversed ordering of the distributions in some narrow region
or in a sparsely populated part of the angle space, as, e.g., spotted in the preshock
domain as a consequence of a small fraction of neutrinos that are backscattered in collisions
with infalling nuclei~\cite{Morinaga+2019}.

Our findings are reminiscent of the individual points of ELN crossings that were identified 
deep inside the PNS by \cite{DelfanAzari+2019} in the 2D simulation of this 11.2\,M$_\odot$ model
and by \cite{Abbar+2019} in the 3D simulations of the same 11.2\,M$_\odot$ progenitor and
of a 27\,M$_\odot$ model. 
In the following section we will argue that the regions of fast-flavor instability
form thin 2D layers in 3D space rather than isolated points, and the identification
of individual points in our analysis is an artifact connected with the discretization of the
physical variables in the numerical treatment. This conclusion does probably also apply
to the results of the previous investigations in Refs.~\cite{DelfanAzari+2019,Abbar+2019}.

\begin{figure*}[!t]
\begin{centering}
\includegraphics[width=2\columnwidth]{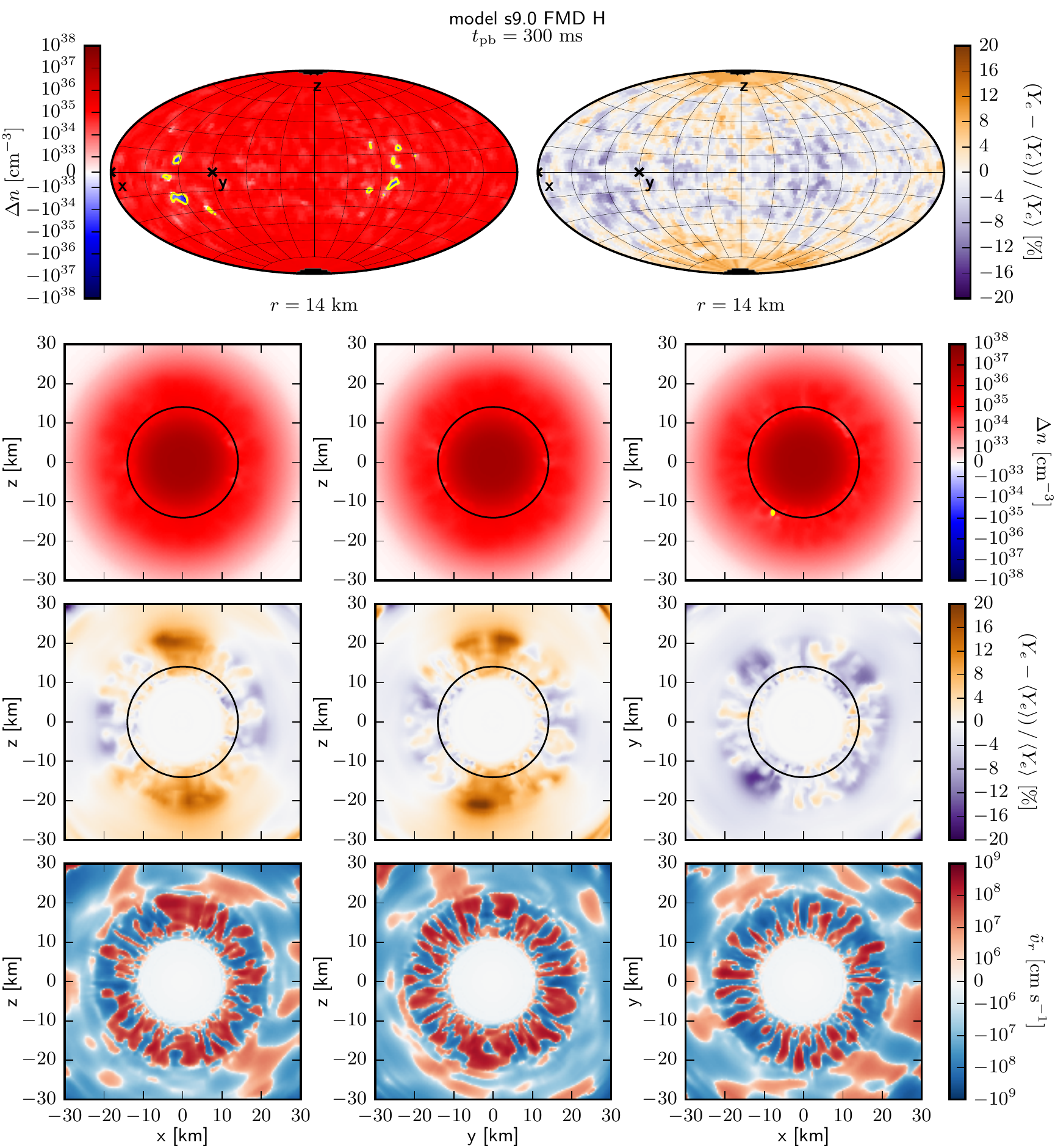}
\end{centering}
\caption{{\em Top row:} Aitoff projections of $\Delta n_\nu = n_{\nu_e}-n_{\bar\nu_e}$ 
({\em left}) and the variations of $Y_e$ relative to the angle-averaged value ({\em right}) 
at a radius of 14\,km inside 
the PNS and post-bounce time of 300\,ms for our 9\,M$_\odot$ model.
The directions of the $x$, $y$, and $z$-axes of the computational
polar grid are denoted by black crosses.
{\em Second row:} Corresponding cross-sectional cuts in the $x$-$z$, $y$-$z$, and 
$x$-$y$-planes with $\Delta n_\nu$ color-coded.
{\em Third row:} Variations of $Y_e$ in these cut planes. 
The radius of $r = 14$\,km is marked by black circles. 
{\em Bottom row:} Radial velocities of the stellar plasma, $\tilde v_r$,
in the cut planes. The convective shell in the PNS is visible by the quasi-regular
pattern of convection cells.
In the plots of $\Delta n_\nu$ red indicates positive values, blue negative 
values, the boundaries between both are locations with $\Delta n_\nu\approx 0$,
where flavor instability, i.e.\ ELN crossings, are expected (highlighted by yellow
lines). The first small raisin-like volumes with $\bar\nu_e$ excess signalled by
negative $\Delta n_\nu$ can be found at locations of particularly low $Y_e$
(intense blue hues for negative $Y_e$ fluctuations relative to the average
value). The spatial variations of $Y_e$ are connected to lepton-rich convective 
updrafts, which carry electron-lepton number from the convectively stable PNS
core outward, and more lepton-poor convective downdrafts. Larger
red patches that mark buoyantly rising plasma in the bottom panels are therefore
correlated with bigger orange regions in the panels of the third row, and more 
extended blue inward flows in the bottom panels coincide with the deepest-blue 
areas in the panels of the third row.
}
\label{fig:9Mregions300ms}
\end{figure*}

\begin{figure*}[!t]
\begin{centering}
\includegraphics[width=2\columnwidth]{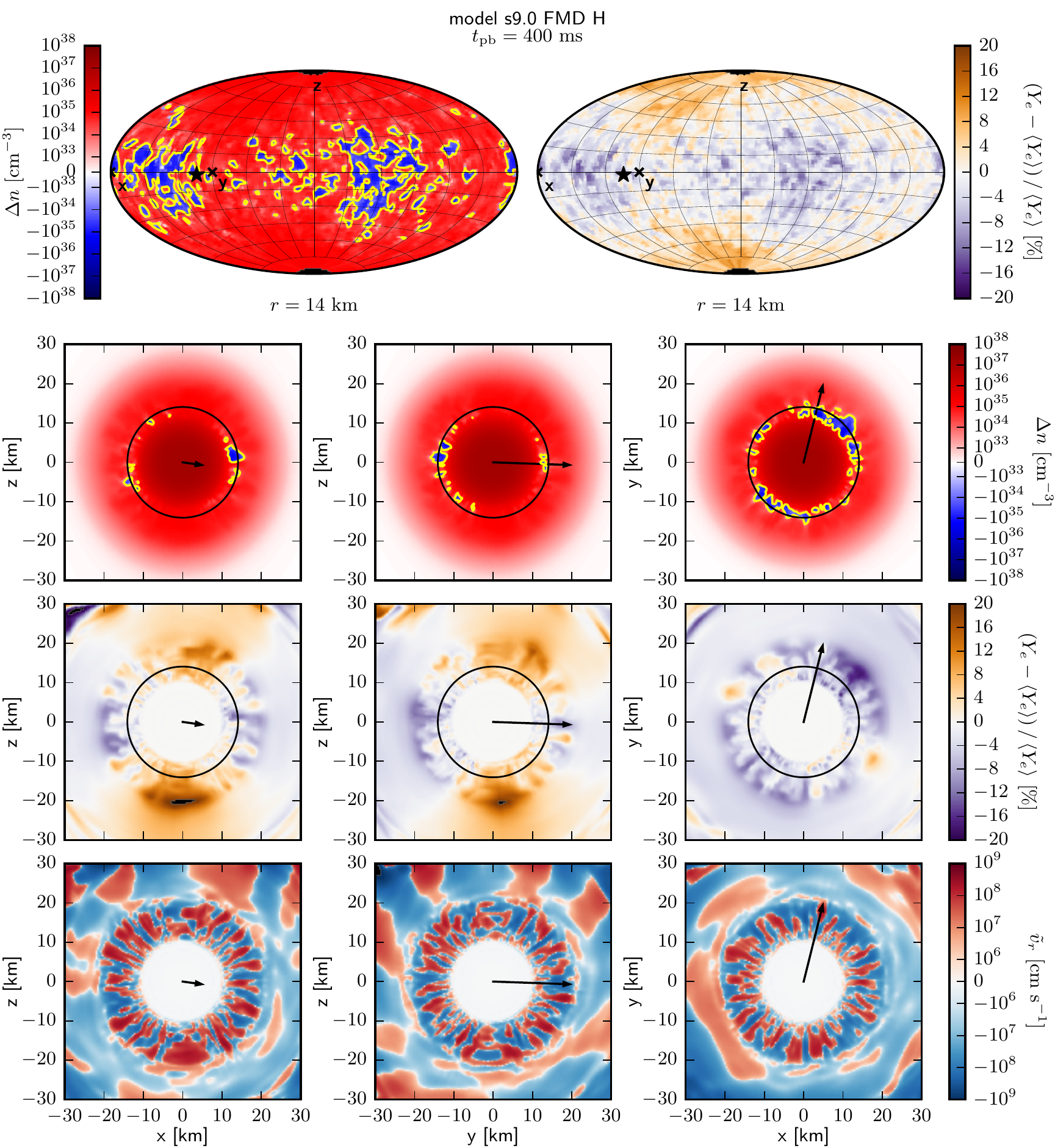}
\end{centering}
\caption{Same as Fig.~\ref{fig:9Mregions300ms}, but at 400\,ms after core bounce.
At this time the LESA lepton-emission dipole has become prominent and its direction
is indicated by a black asterisk on the Aitoff projections and by a black arrow in
the cross-sectional cuts. During all of the model evolution the LESA dipole 
vector direction is close to the $+y$-axis (see Model s9.0\,FMD\,H in figure~3 of 
\cite{Glas:2018vcs}).}
\label{fig:9Mregions400ms}
\end{figure*}

\begin{figure*}[!t]
\begin{centering}
\includegraphics[width=2\columnwidth]{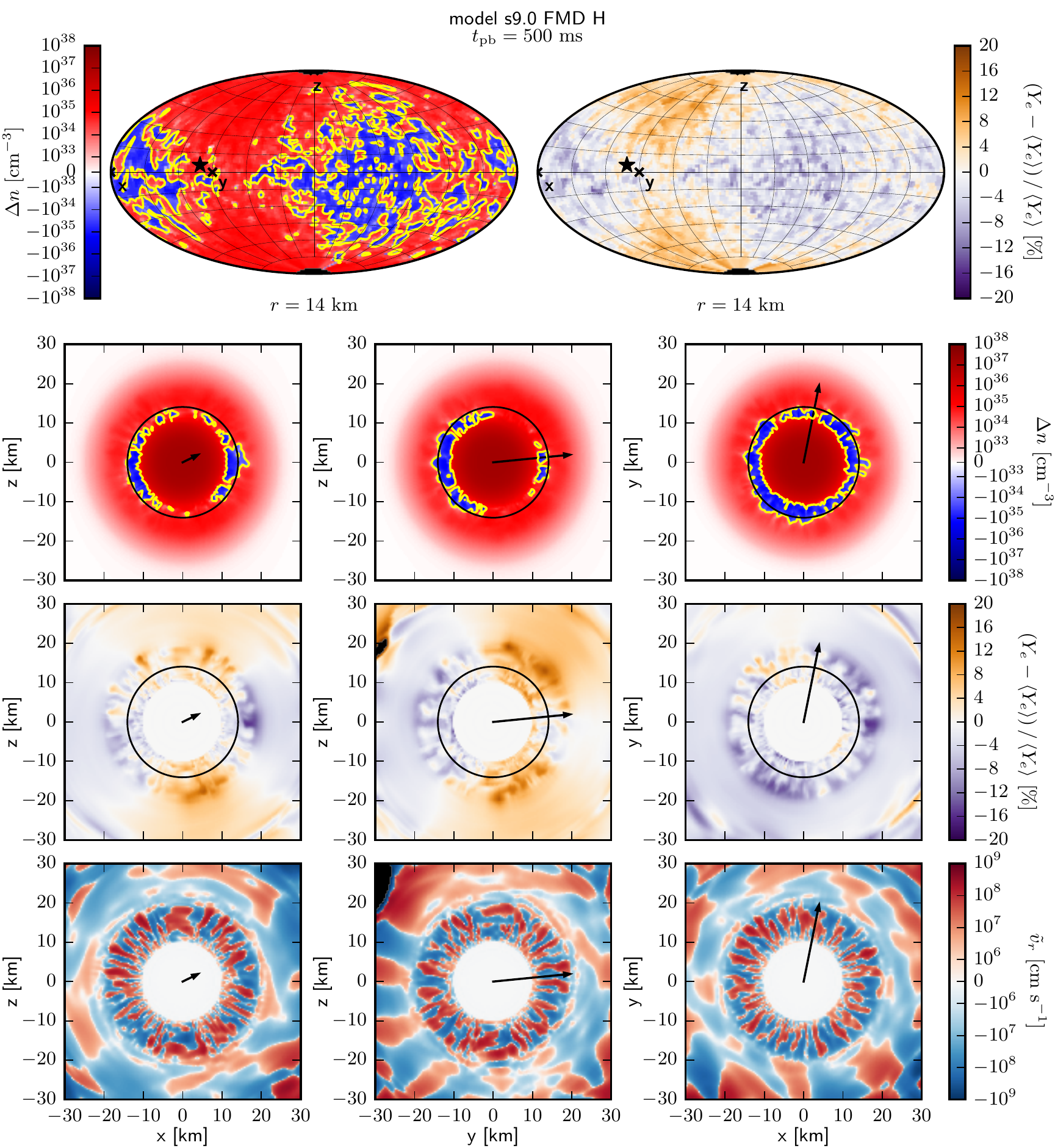}
\end{centering}
\caption{Same as Fig.~\ref{fig:9Mregions300ms}, but at 500\,ms after core bounce.
Regions of negative $\Delta n_\nu$ are more widespread in the anti-LESA direction
where $Y_e$ is lower.}
\label{fig:9Mregions500ms}
\end{figure*}

\begin{figure*}[!t]
\begin{centering}
\includegraphics[width=2\columnwidth]{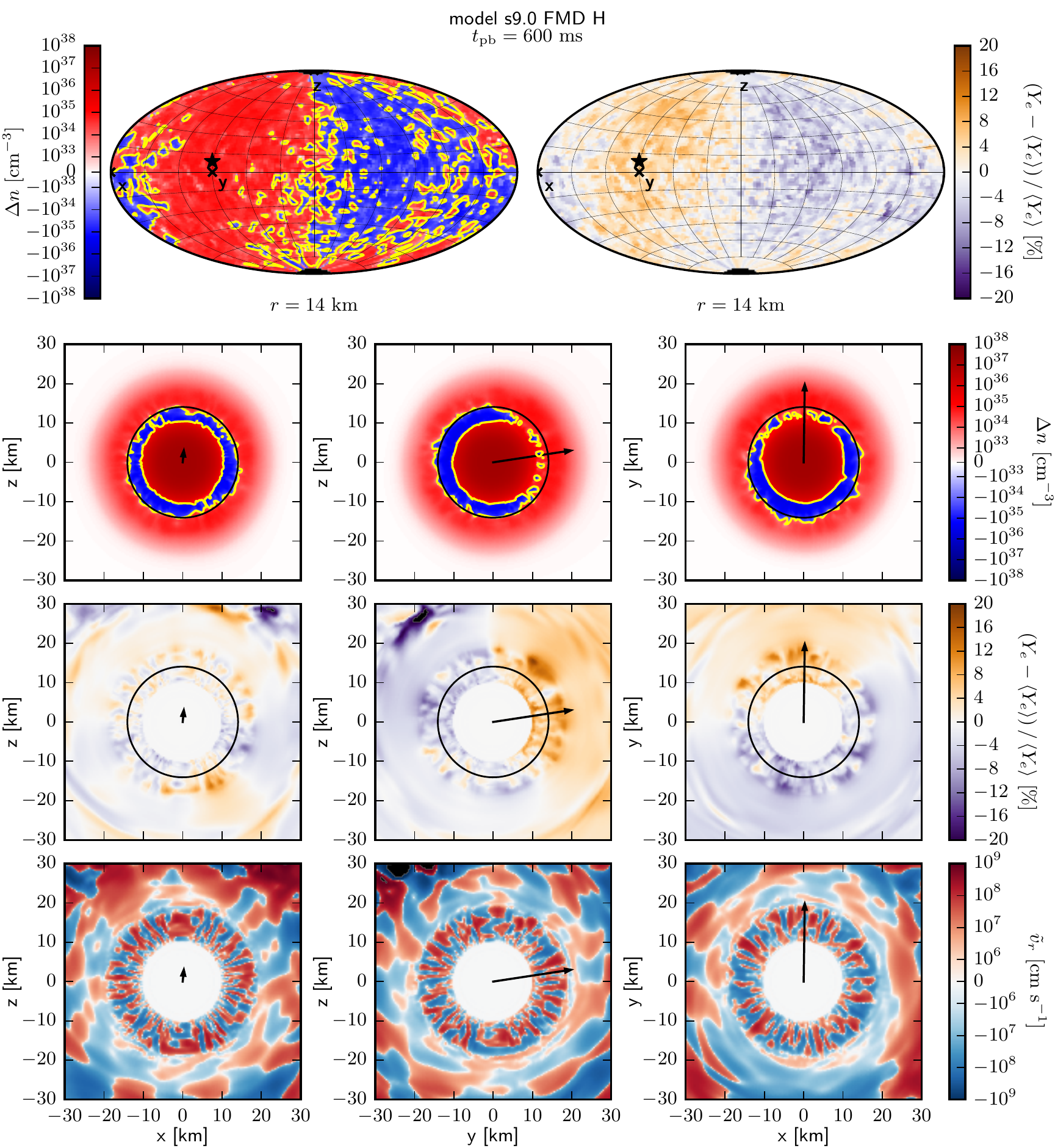}
\end{centering}
\caption{Same as Fig.~\ref{fig:9Mregions300ms}, but at 600\,ms after core bounce.
The LESA dipole is very prominent at this time, corresponding to a clear 
hemispheric asymmetry of $Y_e$ in the convective shell inside the PNS and 
the overlying outer PNS layers. Volumes with $\Delta n_\nu < 0$ and flavor-unstable
boundaries are concentrated mostly in the hemisphere pointing opposite to the
LESA dipole vector (which is indicated by black asterisks and arrows).}
\label{fig:9Mregions600ms}
\end{figure*}

\begin{figure*}[!t]
\begin{centering}
\includegraphics[width=1.01\columnwidth]{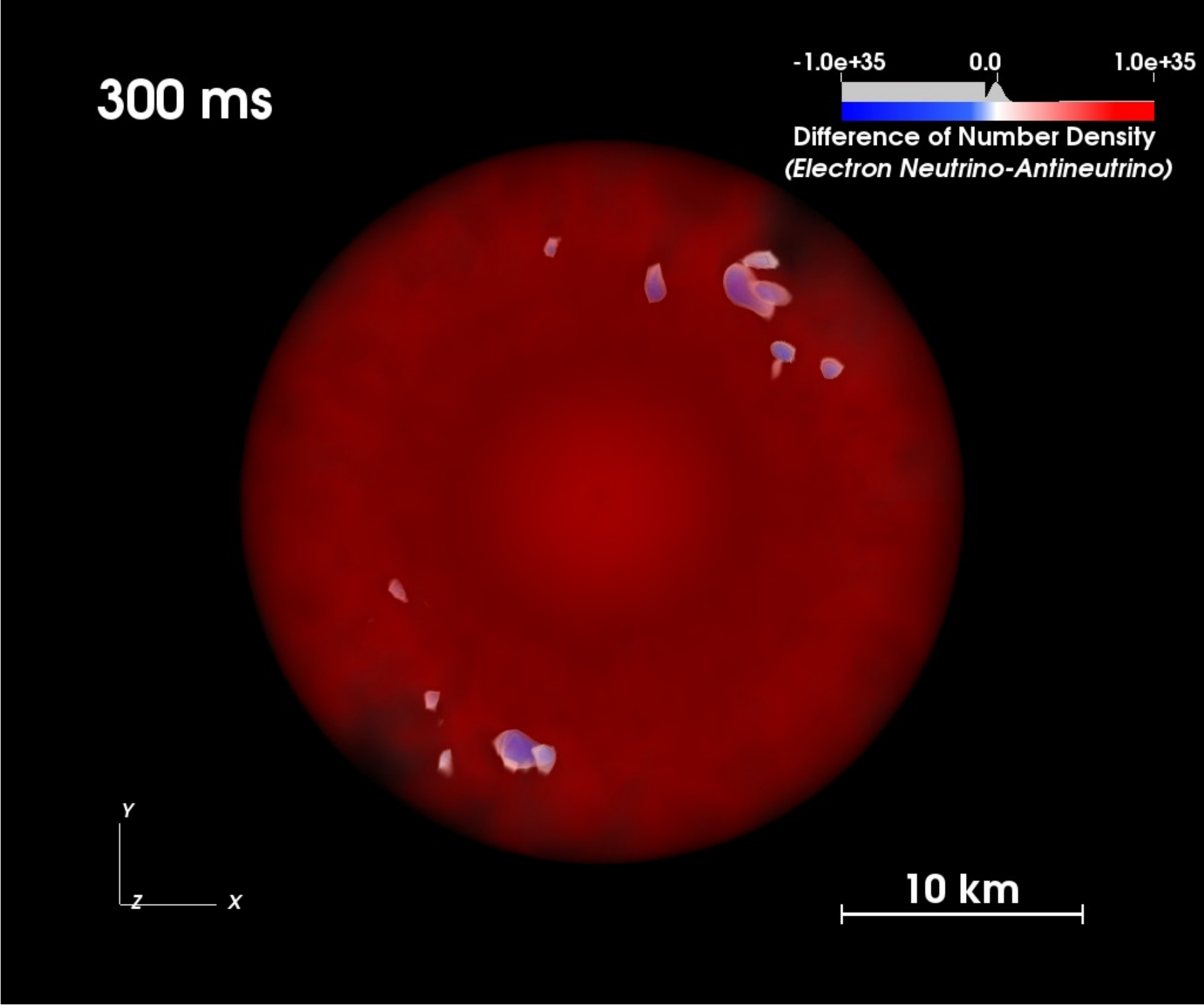}\hskip5pt
\includegraphics[width=1.01\columnwidth]{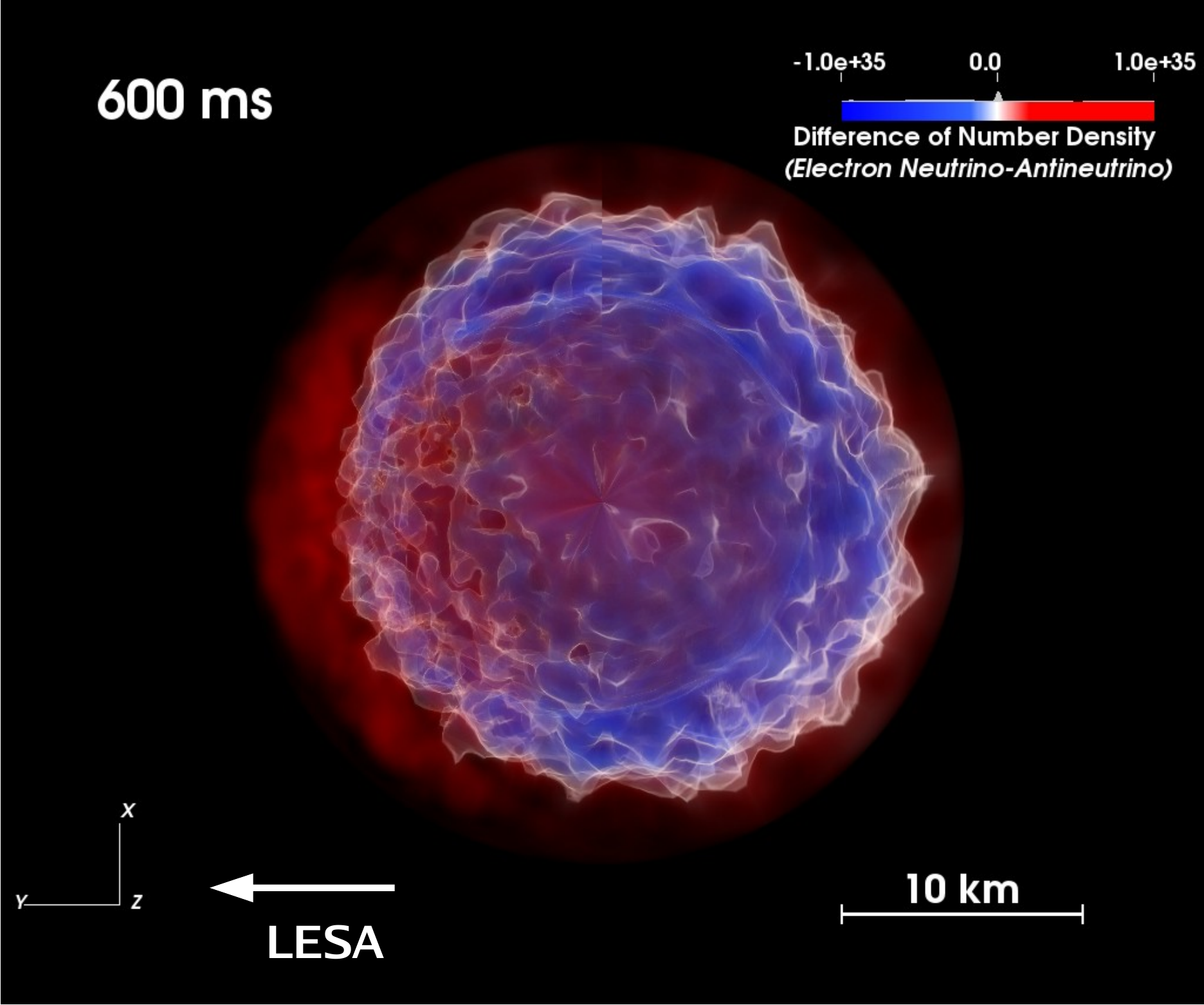}
\end{centering}
\caption{3D volume renderings of $\Delta n_\nu$ in the 9\,M$_\odot$ model at
300\,ms after bounce ({\em left}) and 600\,ms after bounce ({\em right}).
Red hues indicate excess of $\nu_e$ and thus positive $\Delta n_\nu$,
blue hues excess of $\bar\nu_e$ and therefore negative values of 
$\Delta n_\nu$. Flavor-unstable locations are in between near
$\Delta n_\nu = 0$ (whitish). First ``raisins'' with flavor-unstable skins
become visible at about 300\,ms, whereas at 600\,ms flavor-unstable
locations can be found near a radius of 14\,km in the whole
convective layer of the PNS. Note the pronounced hemispheric
asymmetry of the flavor-unstable 2D surface, which is connected
in the anti-LESA direction and more perforated in the hemisphere
which the LESA dipole vector points to (namely the $+y$ direction,
as indicated by the white arrow next to the tripod).}
\label{fig:3Drenderings}
\end{figure*}

\subsection{Conditions for ELN crossings}
\label{sec:3B}

Similar to Refs.~\cite{DelfanAzari+2019} and \cite{Abbar+2019} our points of flavor
instability are located deep inside the PNS, i.e. below the neutrinospheres in a
region where neutrinos diffuse and their phase-space distributions are very close
to those of local chemical equilibrium.

Before we discuss in detail why we ---and probably also the references mentioned 
above--- have found only isolated points instead of extended regions of flavor
instability, we will
introduce a simplified and approximative criterion for the instability, which
makes the underlying physics more transparent. In order to achieve this, we
take advantage of the fact that the nonradial moments are much smaller than the
radial ones in the diffusive core where we find instability. Moreover, in this 
region $V^{rr} = V^{\theta\theta} = V^{\phi\phi}$ holds because the neutrino
phase-space distributions are nearly isotropic. In this limit, which effectively 
corresponds to the 1D case, Eq.~(\ref{eq:disp0}) is explicitly quadratic in 
$\omega\,,$ namely
\begin{equation}
\bigl(\omega+V^{00}\bigr)\bigl(\omega-V^{rr}\bigr)+\bigl(V^{0r}\bigr)^2=0\,\,,
\label{eq:qeq}
\end{equation}
with the solution
\begin{equation}
\omega=\frac{1}{2}\biggl(V^{rr}-V^{00}\pm\sqrt{\left(V^{00}+V^{rr}\right)^2-4\left(V^{0r}\right)^2}\biggr)\,.
\label{eq:dispapprox}
\end{equation}
With the additional relation that $V^{rr}=\frac{1}{3}\,V^{00}$, 
which is valid to high 
accuracy in the diffusion regime,  the condition for instability becomes
\begin{equation}
{\cal F} =\frac{16}{9}\left(V^{00}\right)^2-4\left(V^{0r}\right)^2 <0\,.
\label{eq:discond}
\end{equation}
In terms of the differences of number densities $\Delta n_{\nu}$ and radial
fluxes $\Delta F_{\nu}^r$ of $\nu_e$ and $\bar\nu_e$ (see Eqs.~\ref{eq:equivalence_moments1}
and \ref{eq:equivalence_moments2}) this instability condition reads
\begin{equation}
{\cal F}\,(\sqrt{2} G_F)^{-1} =\frac{16}{9}\left(\Delta n_\nu\right)^2
-\frac{4}{c^2}\left(\Delta F_\nu^r \right)^2 <0\,.
\label{eq:discond2}
\end{equation}
The presence of flavor instability is thus mostly dependent on $V^{00}$ and $V^{0r}$, 
i.e., on the differences of the neutrino number densities and number-flux densities of $\nu_e$ 
and $\bar\nu_e$. We explicitly checked that Eq.~(\ref{eq:dispapprox}) gives almost identical results to what is shown in Fig. \ref{fig:Im_omega_numerical}.

A subtlety concerns the evaluation of the instability condition either with the 
moments in the reference frame comoving with the stellar fluid, where the 
neutrino quantities (i.e., the angular moments) are computed by
the numerical transport code, or in the laboratory frame. In a full-angle
treatment of the neutrino transport this corresponds to the question whether
ELN crossings shall be searched for with the angular distributions of $\nu_e$
and $\bar\nu_e$ in the comoving frame or in the lab frame. We will show here
that at the level of the angular moments, which we use for our analysis,
the results are basically independent of the specific frame where the neutrino
moments are evaluated. 
For fluid velocities $\tilde v \ll c$, the lab-frame and
comoving frame moments are related to lowest order in $\tilde v/c$ through
\begin{eqnarray}
n_\nu^\mathrm{lab} &\approx& n_\nu \,,\label{eq:comov_lab1}\\
F_\nu^{r,\mathrm{lab}} &\approx& F_\nu^r + \tilde v_r n_\nu \,,\label{eq:comov_lab2}\\
P_\nu^{rr,\mathrm{lab}} &\approx& P_\nu^{rr} \,.\label{eq:comov_lab3}
\end{eqnarray}
In the frame transformations of $n_\nu$ (Eq.~\ref{eq:comov_lab1}) and
$P_\nu^{rr}$ (Eq.~\ref{eq:comov_lab3}) we omit terms such as
$\frac{1}{c^2}\,\tilde v_i F_\nu^i$ and $\frac{1}{c^2}\,\tilde v^r F_\nu^r$, respectively. 
Since $\tilde v\ll c$ holds and in the diffusion regime interior to the neutrinospheres
also $\frac{1}{c}\,|F_\nu^i|\ll n_\nu$ 
applies\footnote{This can be easily verified by
comparing the left panels in the top and third rows of Fig.~\ref{fig:moments-radius}
for $r \lesssim 30$\,km and taking into account that $|F_\nu^\theta|\approx|F_\nu^\phi|
\lesssim |F_\nu^r|$.}, 
the disregarded terms are many orders of magnitude smaller than the leading ones
that we retain in Eqs.~(\ref{eq:comov_lab1}) and (\ref{eq:comov_lab3}).
Replacing the comoving-frame quantities in Eq.~(\ref{eq:discond2}) by the 
lab-frame ones of Eqs.~(\ref{eq:comov_lab1})--(\ref{eq:comov_lab3}), we obtain
\begin{eqnarray}
{\cal F}\,(\sqrt{2} G_F)^{-1} &=&\frac{16}{9}\left(\Delta n_\nu^\mathrm{lab}\right)^2
\left(1 - \frac{9}{4}\,\frac{\tilde v_r^2}{c^2}\right) \nonumber \\
&-&\frac{4}{c^2}\,\Delta F_\nu^{r,\mathrm{lab}}\left(\Delta F_\nu^{r,\mathrm{lab}} 
- 2\tilde v_r\Delta n_\nu^\mathrm{lab} \right) \nonumber \\
&<& 0\,.
\label{eq:discond2lab}
\end{eqnarray}
Since $\tilde v_r \ll c$ and, as we shall argue below, the condition can be fulfilled
only when $\Delta n_\nu\approx \Delta n_\nu^\mathrm{lab}\approx 0$, the relation
in Eq.~(\ref{eq:discond2lab}) is basically identical with
\begin{equation}
{\cal F}\,(\sqrt{2} G_F)^{-1} \approx\frac{16}{9}\left(\Delta n_\nu^\mathrm{lab}\right)^2
-\frac{4}{c^2}\,(\Delta F_\nu^{r,\mathrm{lab}})^2 < 0\,, 
\label{eq:discond2.2}
\end{equation}
which is identical to the instability condition of Eq.~(\ref{eq:discond2}).

Figure~\ref{fig:moments-radius} confirms that indeed it does not matter 
whether the analysis is performed with lab-frame or comoving-frame moments for 
the neutrinos. The figure shows, in both reference frames, radial profiles 
of the number densities 
$n_\nu$ of $\nu_e$ and $\bar\nu_e$ individually and their difference 
for the 9\,M$_\odot$ model at a post-bounce time of 300\,ms
(four upper left panels), the corresponding second angular moments $P_\nu^{rr}$
and their difference (four upper right panels), the
radial neutrino-flux densities $F_\nu^r$ and their difference (four lower left
panels) and the ``flavor-instability functional'' ${\cal F}$ of
Eqs.~(\ref{eq:discond2}) and (\ref{eq:discond2.2}) (four lower right panels).
The angular direction ($\theta$, $\phi$) for the radial ray was chosen such that
one of the instability points visible in the top plot of 
Fig.~\ref{fig:Im_omega_numerical}
was crossed. This can be seen in the four panels on the lower right of
Fig.~\ref{fig:moments-radius}, where at $r \approx 14$\,km the flavor-instability
condition is fulfilled. The four upper left panels demonstrate that at this
location $n_{\nu_e}$ and $n_{\bar\nu_e}$ are approximately equal. 
Lab-frame and comoving-frame quantities exhibit exactly the same behavior.

A comparison of the four upper left and four upper right panels shows that
the same conclusion can be drawn from inspection of $P_\nu^{rr}$, because
in the diffusion region $P_\nu^{rr} = \frac{1}{3}\,n_\nu$ is very well  
fulfilled. This relation does not hold any longer when neutrinos begin to
decouple from the stellar medium near the neutrinosphere and undergo the
transition to free streaming outside. In this case $P_\nu^{rr} \to n_\nu$
asymptotically for $r\to \infty$, and therefore our flavor-instability conditions 
of Eqs.~(\ref{eq:discond2}) and (\ref{eq:discond2.2}) are not valid any more.
In the displayed model this is the case for radii $r \gtrsim 30$\,km,
for which reason the negative values of the flavor-instability functional for
$r \gtrsim 40$\,km {\it do not} signal flavor instability in this region
exterior to the PNS.

The two lower right bottom panels of Fig.~\ref{fig:moments-radius}
also display the term $\frac{16}{9}\,(\Delta n_\nu)^2$ as part of the 
flavor-instability functional for comparison with the full expression.
One can see that this term usually dominates the second one, 
$\frac{4}{c^2}\,(\Delta F_\nu^r)^2$, 
by several (typically by 2--3) orders of magnitude. This can also
be directly verified by comparing $\Delta n_\nu$ in the upper
left panels with $\frac{1}{c}\,\Delta F_\nu^r$ displayed in the lower 
left panels. We remark in passing that strongly negative values of the 
$\bar\nu_e$ flux in the comoving frame occur because of a local temperature
maximum that drives the diffusion flux of $\bar\nu_e$ inward while the
more degeneracy-driven diffusion flux of $\nu_e$ can still be outward
directed. Although the lab-frame and
comoving-frame fluxes are considerably different (because the advective
component $v_r\,n_\nu$ can dominate the diffusive component in the 
convection layer of the PNS), the radial profiles of $\Delta F_\nu^r$
are more similar for lab-frame and comoving-frame fluxes, and the  
instability functional ${\cal F}$ in the lower right panels does not
exhibit any visible frame dependence.

There are severe consequences of this huge imbalance between the first 
and the second term in the flavor-instability functional ${\cal F}$
when searching for
ELN crossing points by evaluating the functional with discretized numerical 
results. In order to detect such points, i.e.\ in order to find grid locations
where ${\cal F} < 0$, the term $\frac{16}{9}\,(\Delta n_\nu)^2$ must be very
close to zero at exactly such grid positions, because only then the small 
second term can lead to a negative value of ${\cal F}$. If, however, the 
discrete grid points are too far away from the root of ${\cal F}$, the
values of $\frac{16}{9}\,(\Delta n_\nu)^2$ at these points may be so large 
that the second term $\frac{4}{c^2}\,(\Delta F_\nu^r)^2$ does not achieve
to produce negative values of ${\cal F}$. This, in fact, is likely to happen 
in the far majority of all cases where the physical conditions enable
flavor instability, and only in a minor fraction of such locations the
discretized spatial points of the computational grid coincide incidentally 
with locations where the combination of terms can yield ${\cal F} < 0$,
and thus signal the presence of flavor-unstable conditions.

An example of such a missed point of instability can also be spotted in 
Fig.~\ref{fig:moments-radius}. The plots of ${\cal F}$ in the lower
right panels show two local minima between 10\,km and 20\,km. Only in
the case of the left one the minimum of ${\cal F}$ reaches a negative
value, but for the right one the minimum is still on the positive 
side. Is the condition of flavor instability also fulfilled at this
position and just not detected by the numerical analysis? Indeed, this is 
the situation as visualized in detail by a close-up of the region of
the two local minima of ${\cal F}$ in Fig.~\ref{fig:Instability_criterion}
with crosses marking the positions of all radial mesh points of the
computational grid. The orange line in the two panels corresponds 
to the radial direction chosen for the profiles in 
Fig.~\ref{fig:moments-radius}.
It is obvious that $\Delta n_\nu$ (upper panel of 
Fig.~\ref{fig:Instability_criterion}) has two zero crossings and
becomes negative between these two roots. If a mesh point happens to be 
close to the root, the small, second term in Eqs.~(\ref{eq:discond2})
and (\ref{eq:discond2.2}) achieves to drive
${\cal F}$ (lower panel) to the negative side. Such a situation occurs 
for the left one of the two roots along the radial direction at
$(\theta,\phi) = (107^\circ,57.4^\circ)$ and for the right root in the case 
of $(\theta,\phi) = (111^\circ,57.4^\circ)$. If, however, the mesh points are 
too far away from the root, then ${\cal F}$ remains positive at all discrete
points of the grid. This is the situation for the direction corresponding
to $(\theta,\phi) = (109^\circ,57.4^\circ)$
displayed in Fig.~\ref{fig:Instability_criterion}, although also in this
case $\Delta n_\nu$ possesses two roots. The fourth selected case in this
figure for $(\theta,\phi) = (105^\circ,57.4^\circ)$ does not exhibit any
change in the sign of $\Delta n_\nu$.

Discretization effects are therefore the reason why only very few points
with ELN crossings could be identified by the analysis so far. Consequently,
only individual, isolated points of instability appeared on the three panels 
of Fig.~\ref{fig:Im_omega_numerical}. This problem would become even more 
severe if the resolution of the computational grid used in our SN
simulations had been coarser.

Our case is only an example how numerical discretization effects may
impede the possibility to detect flavor-unstable conditions that occur 
in narrowly delimited spatial regions. A specific condition such as, e.g.,
our instability criteria of Eqs.~(\ref{eq:disp0}), (\ref{eq:discond2}),
and (\ref{eq:discond2.2}) or any other analytical criterion relating physical 
quantities that are available on a discrete numerical mesh, may fail to 
identify the spatial locations of instability. We speculate that also the
investigations in Refs.~\cite{DelfanAzari+2019} and \cite{Abbar+2019},
determining ELN crossings by using angular distributions, suffered from
the finite numerical resolution of the underlying SN simulations and
therefore failed to find more spatial points of flavor-unstable conditions. 

Actually, the problems encountered in our analysis with discretized 
physical variables can easily be circumvented. Evaluating the relations
of Eqs.~(\ref{eq:discond2}) or (\ref{eq:discond2.2}) to search for spatial 
locations where ${\cal F} < 0$ (or for roots of ${\cal F}$) on a mesh of 
discrete points is not a promising strategy.
Instead, it is preferable to look for regions where $\Delta n_\nu$
changes its sign, which we understood as a sufficient
condition to obtain roots of ${\cal F}$.\footnote{Strictly 
speaking, a sign change of $\Delta n_\nu$
is not necessary to get ${\cal F} < 0$, but this condition for ${\cal F}$
can also be fulfilled when $\Delta n_\nu$ dips nearly to zero while still 
remaining positive. However, because $\Delta n_\nu$ and $\frac{1}{c}\,\Delta F_\nu^r$
are orders of magnitude different in the diffusion regime (compare left 
panels in the second and fourth rows of Fig.~\ref{fig:moments-radius}), such
a situation is highly fine-tuned and not as common in hot PNSs as sign 
changes of $\Delta n_\nu$. This is obvious from our analysis of the 
time evolution of PNSs in two 3D SN simulations.}
When $\Delta n_\nu$ changes its
sign between two grid points, there must be a root of this quantity
between the two points. Close to this root ${\cal F}$ will become negative,
unless $\Delta F_\nu^r$ vanishes in this region. In such a pathological
and very rare situation our approximate flavor-instability criterion based 
on a few angular moments of the neutrino phase-space distributions does not
provide conclusive information.

\begin{figure*}[!t]
\begin{centering}
\includegraphics[width=2\columnwidth]{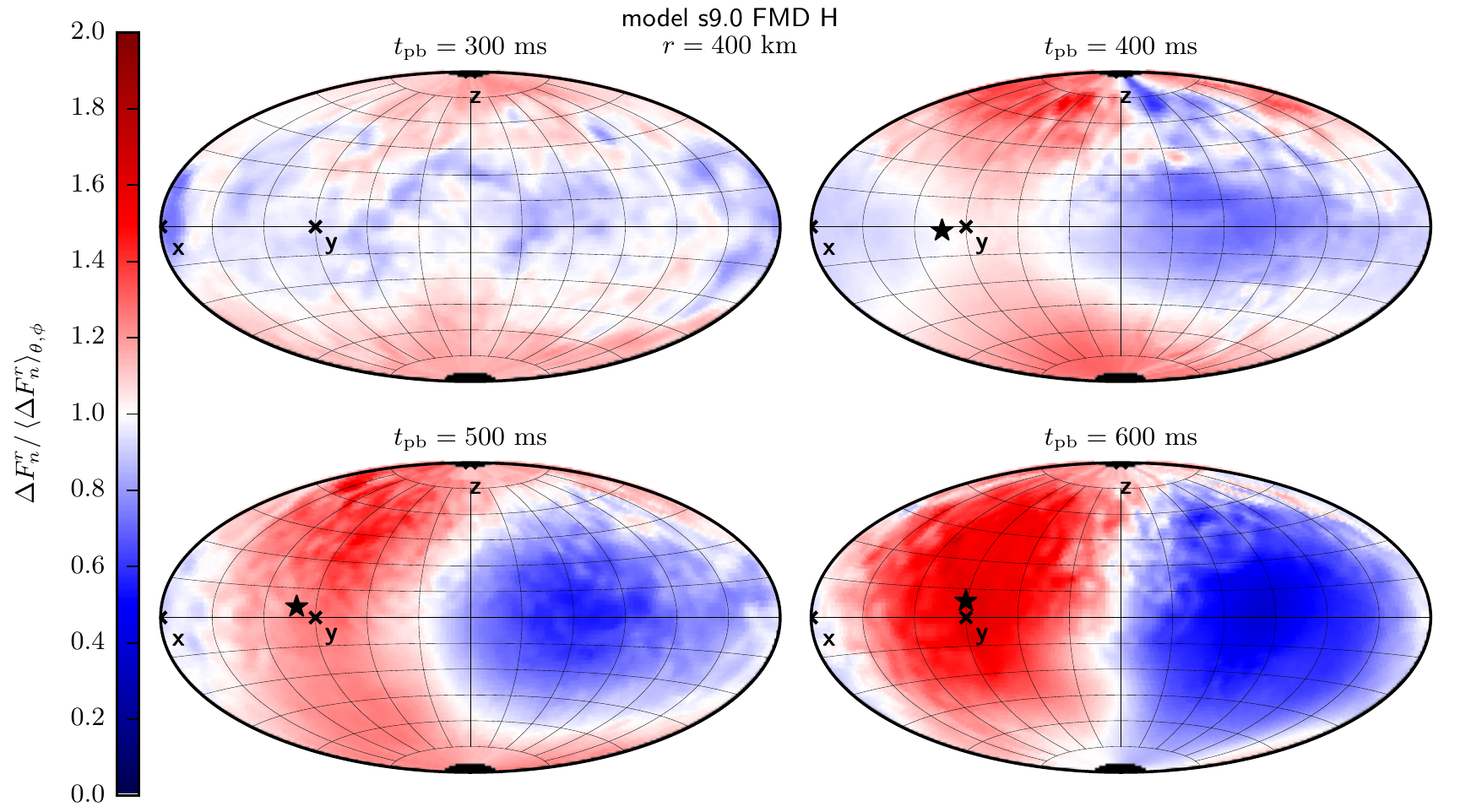}
\end{centering}
\caption{Aitoff projections of the color-coded electron lepton-number
flux, $\Delta F_\nu = F_{\nu_e}-F_{\bar\nu_e}$, normalized to the angle-averaged
value at a radius of 400\,km in the 9\,M$_\odot$ model at 300, 400, 500, and
600\,ms after core bounce ({\em panels from top left to bottom right}).
The development of a pronounced LESA dipole at $t > 300$\,ms is obvious, 
the dipole direction is indicated by black asterisks and is always close to
the $+y$ axis of the computational polar grid (see Model s9.0\,FMD\,H in
\cite{Glas:2018vcs}).}
\label{fig:9M-LESAflux}
\end{figure*}

\begin{figure*}[!t]
\begin{centering}
\includegraphics[width=2\columnwidth]{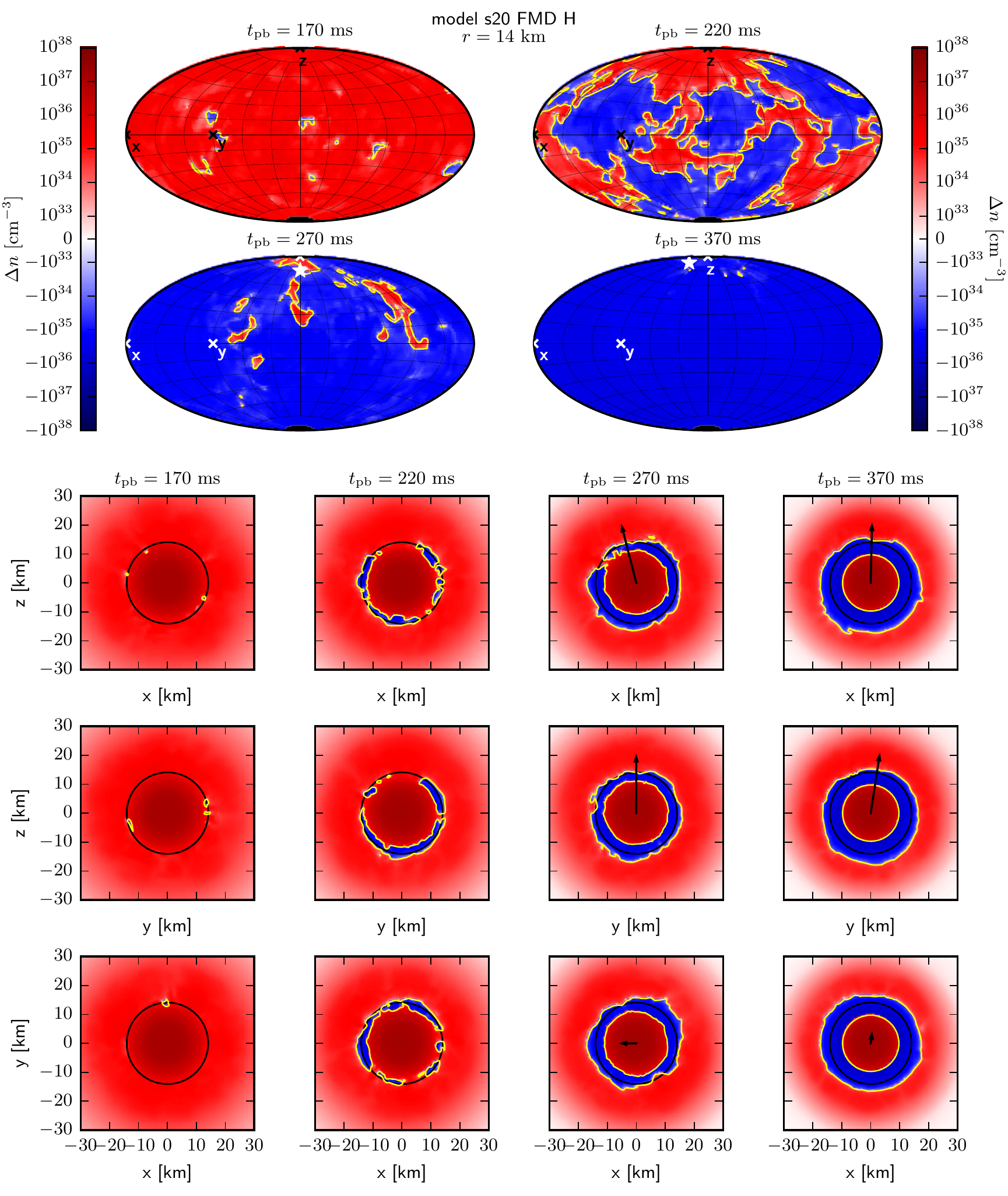}
\end{centering}
\caption{Time evolution of the regions of fast flavor instability
in the 20\,M$_\odot$ model, visualized by Aitoff projections of 
$\Delta n_\nu$ at 14\,km for 170, 220, 270, and 370\,ms after
core bounce. Corresponding cross-sectional cuts are also shown
with $r = 14\,$km indicated by a black circle. At 270\,ms and 370\,ms
a strong LESA dipole has formed (see Model s20\,FMD\,H in
\cite{Glas:2018vcs}); its vector direction is marked
by white asterisks in the Aitoff projections and black arrows in the
cuts. Within only 50\,ms initial ``raisins'' with $\Delta n_\nu < 0$ 
(blue) grow to large ``pancakes'' within a shell around $r = 14$\,km.
At 270\,ms nearly the whole shell is included with a few remaining
``holes'' of $\Delta n_\nu > 0$ (red) remaining in the hemisphere 
of the LESA vector direction. At 370\,ms the shell has become wider
with no remaining holes. Yellow lines indicate the flavor-unstable
locations at the boundaries between regions of $\Delta n_\nu < 0$ and
$\Delta n_\nu > 0$.}
\label{fig:20Mregions170+220ms}
\end{figure*}

\section{Time evolution and physical conditions for ELN crossings}
\label{sec:4}

\subsection{Time evolution}
\label{sec:4A}

With the approximative but numerically robust criterion of 
sign changes of $\Delta n_\nu$, we have evaluated our 9\,M$_\odot$ and
20\,M$_\odot$ models in time to track the evolution of the 
volume of ELN crossings interior to the PNS in our 9\,M$_\odot$ and
20\,M$_\odot$ simulations. 

For the 9\,M$_\odot$ model we show $\Delta n_\nu$ and the
normalized fluctuations of the electron fraction,
$(Y_e - \langle Y_e\rangle)/\langle Y_e\rangle$ (the angle brackets
indicate averages over zenith and azimuthal angles), in full-sphere
Aitoff projections at a radius of
$r = 14$\,km as well as cross-sectional cuts in the $x$-$z$,
$y$-$z$ and $x$-$y$ planes at post-bounce 
times of 300, 400, 500, and 600\,ms in Figs.~\ref{fig:9Mregions300ms},
\ref{fig:9Mregions400ms}, \ref{fig:9Mregions500ms}, and 
\ref{fig:9Mregions600ms}, respectively.

The fluctuations of $Y_e$ are connected to convective updrafts and
downdrafts in the convection layer of the PNS (see
Refs.~\cite{Tamborra:2014aua,Glas:2018vcs}). Convective updrafts
carry electron-lepton number from the convectively stable high-density
core of the PNS outward and therefore exhibit higher values of $Y_e$
than the angle average. In contrast, convective downdrafts are more
lepton-poor, which also means that they contain more neutrons, which
makes them specifically heavier so that they sink inward. The pattern
of $Y_e$ fluctuations mirrors the familiar cell pattern of convection
in spherical shells.

The zero-crossings of $\Delta n_\nu$, and thus the locations very
close to the flavor instability, are highlighted by yellow lines
surrounding the volumes of negative values colored in blue. The
physical thickness of these boundary layers of the $\Delta n_\nu < 0$
volumes,
i.e., the ``skins'' in which the flavor-instability condition is
fulfilled, can be roughly estimated from Eqs.~(\ref{eq:discond2}) or 
(\ref{eq:discond2.2}) by making use of the diffusion approximation
to express the lepton-number flux, $\Delta F_\nu^r = F_{\nu_e}^r - 
F_{\bar\nu_e}^r$ through 
$F_{\nu_i}^r = -D_{\nu_i}\,\partial n_{\nu_i}/\partial r$,
where $D_{\nu_i} = \frac{1}{3}\,c\lambda_{\nu_i}$ is the diffusion coefficient
and $\lambda_{\nu_i}$ the (energy-averaged) mean free path. Introducing 
a mean free path $\bar\lambda$ suitably averaged between
$\nu_e$ and $\bar\nu_e$, we can write for the effective neutrino-lepton
number flux in the diffusion regime: 
\begin{equation}
\Delta F_\nu^r = -\,\frac{1}{3}\,c\bar\lambda\,\,
\frac{\partial(\Delta n_\nu)}{\partial r}\,.
\label{eq:diffusion}
\end{equation}
The requirement for flavor instability, ${\cal F} < 0$ 
(Eqs.~\ref{eq:discond2} and \ref{eq:discond2.2}), implies that 
$|\Delta F_\nu^r| > \frac{2}{3}|\Delta n_\nu|\,c$, or, using 
Eq.~(\ref{eq:diffusion}),
that the radial scale height of changes of the lepton-number density
must be smaller than $\frac{1}{2}\,\bar\lambda$:
\begin{equation}
\left|\Delta n_\nu\,\left(\frac{\partial(\Delta n_\nu)}{\partial r}\right)^{\!-1}
\right| < \frac{1}{2}\,\bar\lambda \,.
\label{eq:skinthickness}
\end{equation}
This means that the thickness of the skins of flavor instability is only
a fraction of the $\nu_e$-$\bar\nu_e$-averaged mean free path $\bar\lambda$,
which is on the order of meters in the PNS region of relevance. 
This skin thickness is one to two orders of magnitude below the
spatial resolution of the best 2D and 3D simulations, which 
explains why a direct evaluation of the flavor-instability
criterion on the discrete grid points of the computational mesh
can find, basically incidentally, only very few locations of ELN 
crossings.

Around 300\,ms first spots of flavor-instability can be seen near the
radius of 14\,km. However, while some moments earlier a small number
of individual, 
isolated points may have fulfilled the instability condition 
${\cal F} < 0$, at 300\,ms these points have already grown to 
2D surfaces enclosing noticeable volumes where $\Delta n_\nu < 0$. 
Such raisin-like inclusions are concentrated around regions where
$Y_e$ is 10--15\% lower than the average. At 400\,ms the 
$\Delta n_\nu < 0$ volumes have considerably grown and partly
merged, enveloped by a coherent surface, besides still existing
smaller droplets. This trend continues until our last displayed
snapshot at 600\,ms. It is obvious that by this time the
quadrupole-dominated pattern of concentrations of instability 
regions that characterizes the situation at 400\,ms and 500\,ms
has evolved to a distribution that is clearly dominated by a 
prominent dipolar asymmetry. While in one hemisphere there are
extended blue 3D regions with $\Delta n_\nu < 0$, the opposite 
hemisphere still exhibits only scattered spots where this condition
is fulfilled. As time goes on these blue regions grow along with
a decreasing electron fraction and rising density because the PNS
gradually deleptonizes and contracts. It is obvious that the
large-scale quadrupolar and dipolar asymmetries of these regions
correlate with such asymmetries in the relative variations of $Y_e$.

Figure~\ref{fig:3Drenderings} visualizes by 3D volume rendering 
the situation in the 9\,M$_\odot$ model at the post-bounce time of
$t = 300$\,ms, when the first scattered ``raisins'' with 
$\Delta n_\nu < 0$ inside and flavor-unstable conditions in their
skins have grown. This is compared to the situation at 600\,ms when
a whole 3D shell between $\sim$10\,km and $\sim$14\,km with
$\bar\nu_e$ excess over $\nu_e$ has developed, more widespread
on one side of the PNS than on the other one. This hemispheric
asymmetry is connected to the LESA phenomenon and establishes
in correlation with the growing dipole amplitude of the LESA,
which can be seen in the Aitoff projections of the normalized 
lepton-number flux, $\Delta F_\nu/\langle\Delta F_\nu\rangle
= (F_{\nu_e} - F_{\bar\nu_e})/\langle\Delta F_\nu\rangle$
(evaluated far outside the PNS at $r = 400$\,km; 
Fig.~\ref{fig:9M-LESAflux}), and of the $Y_e$ distribution
inside the PNS at $r = 14$\,km
(Figs.~\ref{fig:9Mregions300ms}--\ref{fig:9Mregions600ms}).

The black asterisks in the Aitoff plots in all of these figures
and the black arrows in the cross-sectional cuts of 
Figs.~\ref{fig:9Mregions300ms}--\ref{fig:9Mregions600ms} mark
the dipole directions of the lepton-number flux. These markers
are missing in the plots for $t = 300$\,ms, because at that 
early time the emission dipole is not well developed. But once it
is present in a clear way, the dipole direction is extremely stable
in the 9\,M$_\odot$ model, varying only very little around the 
$+y$-direction of the polar coordinate grid of the computation
(see Model s9.0\,FMD\,H in figure~3 of \cite{Glas:2018vcs}). 
It is evident from all these figures that ELN crossings are
less favored on the hemisphere in the direction of the LESA 
dipole vector. As we will discuss in more detail in 
Sec.~\ref{sec:4B}, the reason for this observation is 
the higher electron fraction in the PNS convective shell
on this side. As discussed in \cite{Glas:2018vcs} (see 
also \cite{Tamborra:2014aua}), convection inside the PNS
is stronger in the hemisphere of the LESA dipole direction.
This stronger convection transports electron-lepton number
more efficiently out from the edge of the nonconvective central
core of the PNS, thus raising the $Y_e$ in the convective shell
as well as in the overlying near-surface layers of the PNS
up to the neutrinospheres.  

Figure~\ref{fig:20Mregions170+220ms} presents the evolution
of the flavor-instability regions in our 20\,M$_\odot$ model
by color-coded Aitoff projections and cross-sectional
cuts of $\Delta n_\nu$ in analogy to 
Figs.~\ref{fig:9Mregions300ms}--\ref{fig:9Mregions600ms}.
The time sequence shows the same basic features as visible
in the 9\,$M_\odot$ model, just accelerated because the 
PNS grows faster in mass due to the higher mass-infall rate
in the more massive progenitor. Spots with flavor-unstable
conditions are visible already at 170\,ms after bounce near
a radius of 14\,km and have grown to large coherent 
structures at 220\,ms. At
270\,ms the regions of $\Delta n_\nu < 0$ dominate and 
form a coherent shell around $r = 14$\,km. The very few 
remaining volumes with $\Delta n_\nu > 0$ in this shell have
essentially disappeared until 370\,ms after bounce. 
The direction of the LESA emission
dipole is indicated by white asterisks in the 
Aitoff projections and black arrows in the cross-sectional
cuts at the times when a strong dipole of the lepton-number
emission exists. Again one can notice that flavor-unstable
conditions are more widespread in the anti-LESA direction,
where the electron fraction is lower than in the hemisphere
the LESA dipole vector is pointing to. Accordingly, the last
islands with $\nu_e$ excess (i.e., $\Delta n_\nu > 0$) can
be found on this side of the PNS (see panels for $t = 270$\,ms
in Fig.~\ref{fig:20Mregions170+220ms}).

In the following section we will explain the physical
connection between the conditions of flavor instability and
the evolution of electron fraction $Y_e$, density $\rho$,
and temperature inside the PNS in more detail.

\begin{figure*}[!t]
\begin{centering}
\includegraphics[width=2\columnwidth]{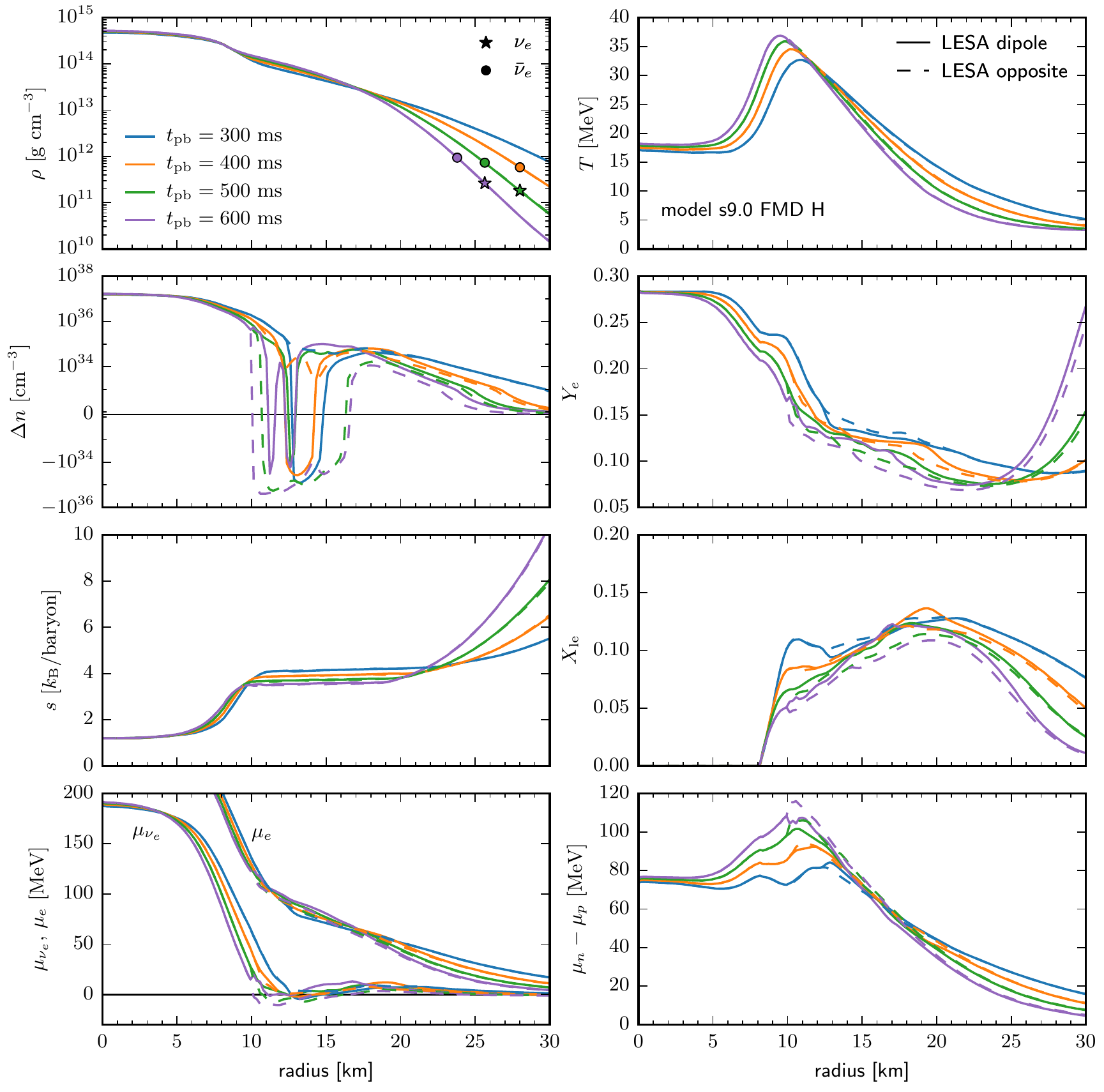}
\end{centering}
\caption{Radial profiles of mass-density ($\rho$; {\em top left}),
temperature ($T$; {\em top right}), difference of $\nu_e$ and $\bar\nu_e$
number densities ($\Delta n_\nu$; {\em second row left}), electron fraction
($Y_e$; {\em second row right}), gas entropy per nucleon ($s$; {\em third row left}),
mass fraction of light nuclei, including $\alpha$ particles, that exist besides
free neutrons and protons ($X_\mathrm{le}$; {\em third row right}),
chemical potentials of electrons and electron
neutrinos ($\mu_e$, $\mu_{\nu_e}$; {\em bottom left})
and chemical potential difference between neutrons and protons
($\mu_n-\mu_p$; {\em bottom right}) in our 9\,M$_\odot$ model at
300, 400, 500, and 600\,ms after core bounce. Solid lines display the
profiles in the direction of the LESA dipole, dashed lines in the opposite
direction. At 300\,ms, when the LESA dipole is still very weak, the later
direction of the LESA dipole vector is chosen, because it stably remains
in the close vicinity of the $+y$-axis of the computational polar grid
once it has developed in this model (see Model s9.0\,FMD\,H in
figure~3 of \cite{Glas:2018vcs}). On the density profiles the locations of
the neutrinospheres of $\nu_e$ and $\bar\nu_e$ (defined by a spectrally
averaged total optical depth of 1) are marked by an asterisk or bullet,
respectively. The convective shell inside the PNS can be recognized by the
region where the entropy profile is flat.}
\label{fig:9Moverview-all-times}
\end{figure*}

\begin{figure}[!]
\begin{centering}
\includegraphics[width=\columnwidth]{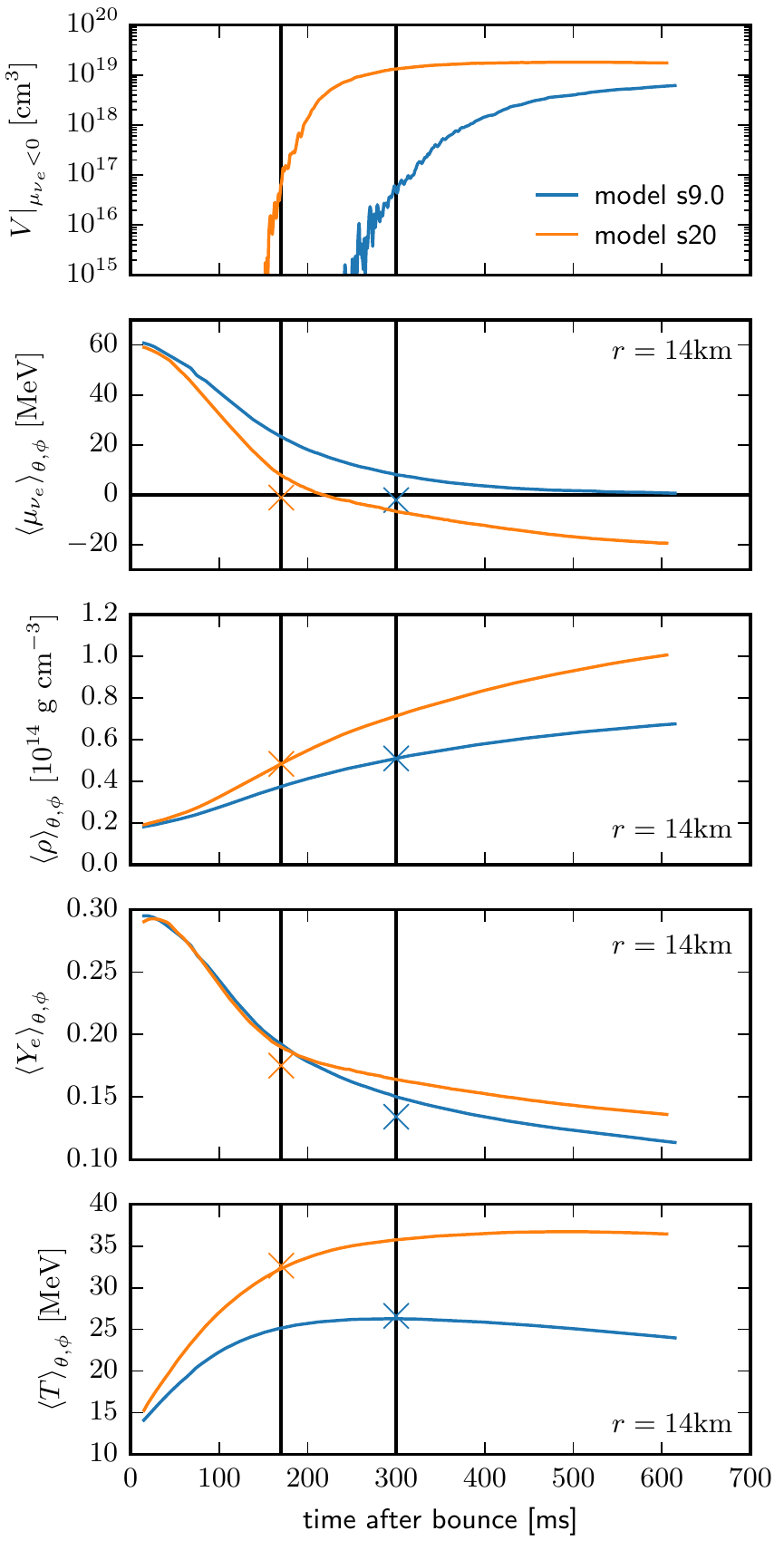}
\end{centering}
\caption{Time-dependent comparison of the 9\,M$_\odot$ and 20\,M$_\odot$ models
with respect to the development of flavor-unstable skins around regions of
$\Delta n_\nu < 0$, measured by the spatial volumes 
$\left.V\right|_{\mu_{\nu_e} < 0}$
enclosed by these boundary layers ({\em top panel}).
Also displayed is the evolution of quantitities that are relevant for
understanding this effect: electon-neutrino chemical potential
($\mu_{\nu_e}$, {\em second panel}), mass density ($\rho$; {\rm third panel}),
electron fraction ($Y_e$; {\em fourth panel}), and gas temperature ($T$;
{\em bottom panel}), all angle-averaged at our representative radius of
$r = 14$\,km. The two vertical black lines mark the earliest moments of 
the evolution of both models (300\,ms and 170\,ms after bounce, respectively)
shown in Figs.~\ref{fig:9Mregions300ms} and \ref{fig:20Mregions170+220ms}.
The crosses indicate the conditions that are found inside the 
raisin-like volumes with $\Delta n_\nu < 0$ at these early times.
Some of these conditions tend to be extreme compared to the 
angular averages, for which reason the crosses do not lie on the curves.
}
\label{fig:compare9+20Msun}
\end{figure}

\begin{figure}[!]
\begin{centering}
\includegraphics[width=\columnwidth]{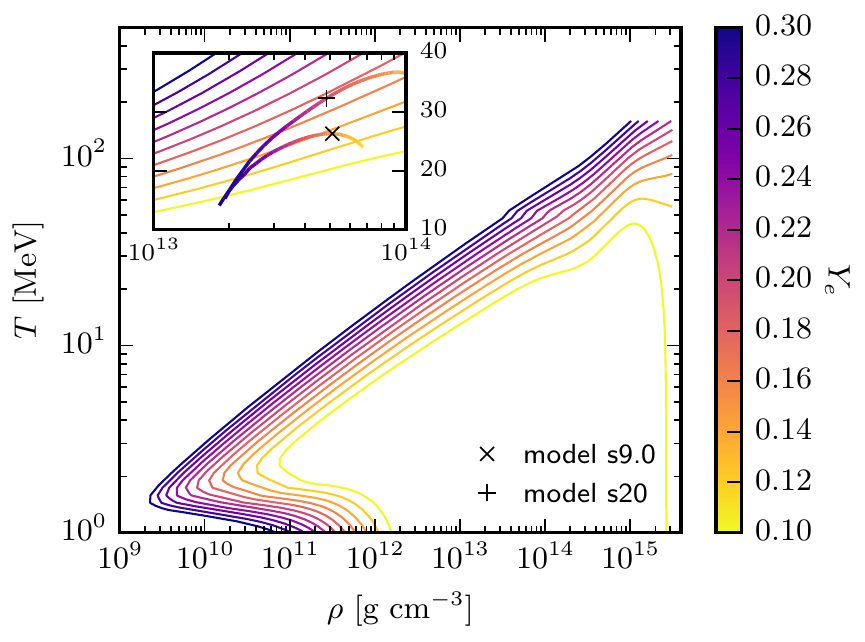}
\end{centering}
\caption{Contours of vanishing neutrino chemical potential
($\mu_{\nu_e} = 0$) in the temperature-density plane for different
constant values of $Y_e$ (color coded), evaluated for the nuclear 
equation of state employed in our 3D SN simulations (SFHo from
\cite{Steiner+2013}). The inset 
shows a zoom of the density interval between $10^{13}$\,g\,cm$^{-3}$
and $10^{14}$\,g\,cm$^{-3}$ with the evolution paths of the 9\,M$_\odot$
and 20\,M$_\odot$ models superimposed, also color-coded for the 
evolving value of $\langle Y_e\rangle$, angle-averaged at a radius 
of 14\,km in each model. Along the tracks $\langle Y_e\rangle$ 
decreases as the PNS
deleptonizes; the evolution therefore proceeds from the lower left
to the upper right. Spatial locations of flavor-instability begin to
appear when $\langle Y_e\rangle$ on the track becomes equal to the $Y_e$
of a $\mu_{\nu_e} = 0$ contour, which is visible by the color of
the evolution track matching the color of the contour. The corresponding 
instants are approximately marked by symbols, corresponding to the times
of the vertical lines in Fig.~\ref{fig:compare9+20Msun} (the cross
belongs to the 9\,M$_\odot$ model, the plus to the 20\,M$_\odot$
simulation).}
\label{fig:sfho}
\end{figure}

\subsection{Properties of ELN crossing points}
\label{sec:4B}

Naturally, the relation $\Delta n_\nu = n_{\nu_e} - n_{\bar\nu_e} 
\approx 0$, which we recognized
in Sec.~\ref{sec:3B} as a necessary condition for ELN crossings in
a regime where neutrinos diffuse and are close to local chemical 
equilibrium, implies that $\mu_{\nu_e}=\mu_e+\mu_p-\mu_n\approx 0$
holds for the electron-neutrino chemical as a function of the 
chemical potentials of electrons, protons, and neutrons. Only then
$\nu_e$ and $\bar\nu_e$ in local chemical equilibrium can have nearly
equal number densities. This was
pointed out already in Refs.~\cite{DelfanAzari+2019} and \cite{Abbar+2019}.

But what is the reason why fast flavor unstable conditions develop
in an increasing volume in the new-born NS? 
The relation of the chemical potentials that determines $\mu_{\nu_e}$ 
suggests that low $Y_e$ and correspondingly low $\mu_e$ might allow
$\mu_{\nu_e}$ to drop to zero and even to negative values.

Figure~\ref{fig:9Moverview-all-times} displays
radial profiles of these and other quantities in the LESA and anti-LESA
directions of the 9\,M$_\odot$ model at four post-bounce times.
Inspecting the different panels shows that the locations where
fast flavor instability is possible ($\Delta n_\nu \approx 0$)
lie within the convective shell in the interior of the
PNS; the convective layer can be recognized by the convectively 
flattened gradient of the entropy per nucleon, $s$. However, neither $Y_e$
nor $\mu_e$ reach a local minimum where the electron-neutrino chemical
potential develops a trough near zero or below, and also the 
abundances of $\alpha$ particles or light elements up to $^4$He are
not particularly abundant in general in the regions where ELN crossings
are favored, in contradiction to arguments in \cite{DelfanAzari+2019}. 
This can be directly concluded from a comparison of the LESA and anti-LESA
directions in Fig.~\ref{fig:9Moverview-all-times}. In the hemisphere
opposite to the LESA dipole vector, $Y_e$ as well as the abundances
of $\alpha$ particles and light elements are lower than in the hemisphere
the LESA vector points to. Nevertheless, flavor unstable locations are more 
widespread in the anti-LESA direction at late times (500, 600\,ms after 
bounce) when the dipole is strong (as discussed in Sec.~\ref{sec:4A}).
Lower values of $Y_e$ enhance the possibility of flavor instability, but 
they are not causal in the first place.

Instead of being linked to peculiarities in the $Y_e$ profile, the 
$\mu_{\nu_e}$ trough is found in a region where $\hat\mu\equiv\mu_n-\mu_p$ 
possesses a local maximum, because high values of $\hat\mu$ reduce
$\mu_{\nu_e}=\mu_e-\hat\mu$ compared to $\mu_e$. 
The local maximum of $\hat\mu$ 
again correlates tightly with a local temperature maximum, which is
a relic of shock heating by the initial propagation of the strong 
shock front formed at the moment of core bounce. The close connection
between maxima of $\hat\mu$ and $T$ can be easily understood from 
considering nondegenerate, nonrelativistic and noninteracting neutrons
and protons as an approximation of nucleons in the subnuclear regime.
For such Boltzmann gases the chemical potential of particle species
$i = n,\,p$ with number density $n_i$ and rest-mass $m_i$ is given by 
$\mu_i = m_ic^2 + k_\mathrm{B}T\,\ln(\Lambda_{\mathrm{th},i}^3n_i/2)$,
where $\Lambda_{\mathrm{th},i} = h\,(2\pi m_i k_\mathrm{B}T)^{-1/2}$
is the thermal wavelength, $k_\mathrm{B}$ the Boltzmann constant, and
$h$ the Planck constant. With $Q \equiv (m_n-m_p)c^2$ one therefore gets:
\begin{equation}
\hat\mu \equiv \mu_n-\mu_p = Q + k_\mathrm{B}T\,\ln\left[\frac{n_n}{n_p}\,\,
\left(\frac{m_p}{m_n}\right)^{3/2}\right]\,.
\label{eq:muhat}
\end{equation}
This relation explains the direct dependence of $\hat\mu$ on the plasma
temperature, and it also implies that lower $Y_e$ in the hemisphere 
opposite to the LESA vector reduce $\mu_{\nu_e}$ not only through lower
values of $\mu_e$ but also through higher ratios of 
$n_n/n_p$.\footnote{We note that the logarithmic dependence
of $\hat\mu$ on $n_n/n_p$ is weak, but also $\mu_e$ depends logarithmically
on $n_e$ when electrons are nondegenerate, and $\mu_e$ is proportional to 
$n_e^{1/3}$ when electrons are extremely degenerate.} 
Non-negligible effects due to nucleon interactions in the density regime
of interest between some $10^{13}$\,g\,cm$^{-3}$ and 
$\sim$$10^{14}$\,g\,cm$^{-3}$ may lead to quantitative changes but 
do not qualitatively affect this argument.

The development of regions with negative electron-neutrino chemical
potential, $\mu_{\nu_e} < 0$, inside the PNS does not only depend on 
temperature and $Y_e$ but also on density. This can be concluded from
Fig.~\ref{fig:compare9+20Msun}, where the evolution of the 9\,M$_\odot$
and 20\,M$_\odot$ models is compared. The upper panel displays the 
growth of the volume of regions with $\mu_{\nu_e} < 0$. Vertical black
lines mark the two earliest instants displayed for both simulations
in Fig.~\ref{fig:9Mregions300ms} and Fig.~\ref{fig:20Mregions170+220ms},
respectively. The solid lines represent angular averages 
of the quantities over the whole
sphere at $r=14$\,km, and the crosses indicate the conditions inside
the (still very small) volumes that fulfill $\Delta n_\nu < 0$ at 300\,ms
(170\,ms) in the 9\,M$_\odot$ (20\,M$_\odot$) model (the two instants
marked by the vertical black lines). Since the conditions in the
$\mu_{\nu_e} < 0$ volumes are special and for some quantities tend to
be extreme compared to the conditions on the rest of the sphere, the
crosses do not lie on the curves of the angular 
averages.\footnote{The crosses correspond to data at
$r=14$\,km and $(\theta,\phi)=(109.00^\circ,57.375^\circ)$ for the 
9\,M$_\odot$ model and  $(\theta,\phi)=(75.00^\circ,86.625^\circ)$
for the 20\,M$_\odot$ case.}

The crosses in the second panel of Fig.~\ref{fig:compare9+20Msun} show
that at the marked instants the electron-neutrino chemical potential 
begins to become negative in the raisin-like spots that are visible
in the Aitoff projections of the two models (Figs.~\ref{fig:9Mregions300ms} 
and \ref{fig:20Mregions170+220ms}), whereas the angle-averaged values
of $\mu_{\nu_e}$ at $r=14$\,km
are still somewhat higher, with a monotonically decreasing trend with 
time. The remaining three panels contain the evolution of density, 
electron fraction, and temperature, again angle-averaged at $r=14$\,km.
In both models flavor-unstable raisin-skins begin to occur when the
density reaches roughly $5\times 10^{13}$\,g\,cm$^{-3}$, whereas $Y_e$ 
and temperature are different between the two models at this time. 

From Fig.~\ref{fig:sfho} it becomes clear where these differences 
originate from. The plot shows a set of isocontours for different,
fixed values of the electron fraction (color-coded) in the 
density-temperature plane. All contours are defined by the condition
$\mu_{\nu_e} = 0$. At each point $(\rho,T)$ this condition is 
fulfilled only for a single value of the electron fraction. 
Each $\mu_{\nu_e} = 0$ contour encloses a $\rho$-$T$-domain
where $\mu_{\nu_e} > 0$ for the $Y_e$ value of the contour, whereas 
$\mu_{\nu_e} < 0$ holds for the same $Y_e$ value outside of this 
$\mu_{\nu_e} = 0$ 
contour (see also figures~11 and 12 in Ref.~\cite{Ruffert+1997}).

In the inset of Fig.~\ref{fig:sfho},
the evolution tracks of both SN models in the
$\rho$-$T$-$Y_e$-space, with all quantities angle-averaged at 
$r = 14$\,km, are superimposed on the contours of vanishing neutrino
chemical potential. The tracks start at low values of temperature and 
density with a high value of the electron fraction, and evolve from
the lower left to the upper right of the inset as the PNS contracts
and heats up by compression and conversion of electron degeneracy
energy to thermal energy (Joule heating). 
ELN crossings become possible when the
electron fraction on the track matches the $Y_e$ value of one of the
contours, i.e. when the color of the track and of a contour are the
same. The symbols mark the $(\rho,T)$ locations when first noticeable 
flavor-unstable spots occur in the two SN runs at $r = 14$\,km
(corresponding to the two vertical lines in Fig.~\ref{fig:compare9+20Msun}). 

In the 20\,M$_\odot$ model the mass of the PNS grows faster than in
the 9\,M$_\odot$ simulation because of the bigger mass-accretion rate
in the more massive progenitor. Therefore the temperature in the PNS
interior increases more rapidly and more steeply with density,
and reaches greater
values, allowing for a match of the electron fraction with one of the
$\mu_{\nu_e} = 0$ isocontours more quickly and at a higher value
of $Y_e$. As a consequence,
fast flavor unstable conditions occur earlier in the 20\,M$_\odot$ model
than in the 9\,M$_\odot$ case.

Compressional and Joule heating and continuous deleptonization 
are characteristic features of the neutrino-cooling evolution of 
new-born NSs. Our study includes a low-mass NS of a 9\,M$_\odot$
progenitor as well as a more massive NS in the 20\,M$_\odot$
model. Both of them develop flavor-unstable conditions in an 
increasing volume of the convective layer (Fig.~\ref{fig:compare9+20Msun}), 
and the rising temperature naturally leads to a match of the decreasing 
$Y_e$ values along
evolution tracks with the $Y_e$ of isocontours for
$\mu_{\nu_e} = 0$ (Fig.~\ref{fig:sfho}). We therefore expect that
fast flavor unstable regions in the deep interior of PNSs are a 
generic phenomenon during PNS cooling.

\section{Conclusions}
\label{sec:5}

We performed a detailed investigation of 3D state-of-the-art SN models
for the presence of fast neutrino flavor instability and to study the
favorable conditions. These fast conversions are associated with crossings
in the angular distribution of the ELN. However, the simulations we 
analyzed did not provide the detailed neutrino angular distributions. 
In order to overcome this limitation, we adopted a novel method for
calculating the growth rate of the instability, which is based only
on the angular moments of the ELN up to second order \cite{Dasgupta:2018ulw}. 
We applied our analysis to SN simulations of 9\,M$_\odot$ and 
20\,M$_{\odot}$ progenitors, which were recently conducted by the
Garching group \cite{Glas:2018oyz,Glas:2018vcs}. In both models we
found conditions for fast flavor
instability deep inside the PNS in a radial range of 
$10\,\mathrm{km}\lesssim r \lesssim 20\,$km, where all flavors of neutrinos 
are in the diffusive regime and close to local chemical equilibrium.

We thus confirm similar recent detections of locations of ELN
crossings in the PNS interior based on Boltzmann transport results
in 2D time-dependent and 2D/3D fixed-background SN models in 
Refs.~\cite{DelfanAzari+2019,Abbar+2019}. However, we showed that the
direct evaluation of flavor-instability conditions with the discretized
output of numerical simulations leads to the identification of only
few individual, isolated points of ELN crossings. We argued that this 
finding on grounds of our method, and most likely also by the
approaches used in the previous analyses, is a numerical artifact 
and misses the full phenomenology of the physical effects.

In reality the spatial locations of fast flavor instability are
extended, narrow boundary layers surrounding volumes in 3D space
where the $\bar\nu_e$ number density exceeds the $\nu_e$ number
density. These surface layers have a thickness of roughly a 
neutrino mean free path and they grow from the skins of initially
scattered, raisin-like inclusions to the envelopes of increasingly
larger regions, until they finally form the inner and outer surfaces
of a closed layer with $n_{\bar\nu_e} > n_{\nu_e}$. 

The region where this happens is located within the convective layer
of the PNS where on the one hand convective transport of lepton 
number causes a rapid decline of the electron fraction, and on the
other hand a local temperature maximum is further increased when
the PNS contracts and compression as well as the conversion of 
electron-degeneracy energy to thermal energy heat the stellar
plasma. Rising density and temperature combined with a decreasing
electron fraction naturally drive this layer in the PNS towards
conditions where the electron-neutrino chemical potential is
$\mu_{\nu_e}\le 0$, implying an excess of $\bar\nu_e$ relative to
$\nu_e$. Since the lepton-number emission dipole associated
with the LESA phenomenon is caused by stronger PNS convection in 
the hemisphere of the dipole direction, 
conditions for ELN crossings are more widespread
in the opposite hemisphere, where PNS convection is weaker and 
therefore the electron fraction is not efficiently replenished by
leptons carried outward from the edge of the nonconvective central
core.

Our result backs the findings in \cite{DelfanAzari+2019,Abbar+2019}
and points to ELN crossings in the PNS convection layer as a generic 
phenomenon during the cooling evolution of new-born NSs. In contrast
to these previous works we have demonstrated that the regions of
fast flavor unstable conditions are not fluctuating and point-like, 
but instead they are large-scale and long-lasting spatial structures.

This opens new directions of research. Presently, however, it is unclear
whether fast flavor conversions in the deep interior of the PNS can
have major consequences for the PNS cooling and/or SN evolution. 
The instability takes place in an extremely thin layer where 
$\mu_{\nu_e}$ is close to zero. Since these locations are deep
inside the neutrinospheres of all neutrino species, neutrinos of
all flavors are very close to chemical equilibrium. Therefore 
$\mu_{\nu_e}\approx 0$ implies that $\nu_e$ and $\bar\nu_e$ possess 
phase-space distributions that are very similar not only to 
each other but also to those of muon and tau neutrinos.
At such conditions flavor exchange might have little
impact on the overall conditions. Given the importance of potential
effects, our work should stimulate further investigations. In view
of the mutual support between our results and those obtained in
Refs.~\cite{DelfanAzari+2019,Abbar+2019}, we conclude that our
method based on angular moments is well suitable to analyze SN
models computed with neutrino transport schemes that do not provide
the detailed neutrino angular distributions.

The presence of fast conversions adds an additional layer of 
complexity to SN simulations.
The current paradigm of flavor conversions is based on a separation 
from the treatment of neutrino interactions and transport.
However, if fast conversions occur in the region of neutrino spectra
formation or even deep inside the PNS, a self-consistent characterization
of both of these phenomena will require new strategies for a simultaneous
treatment of flavor oscillations and collisional effects.
This will be a formidable task, since fast flavor conversions occur 
on time and length scales much shorter than usually resolved in 
global simulations of stellar collapse and explosions. Therefore,
new methodical approaches will be needed. It is obvious that a huge
amount of work still remains to be done to understand the role of neutrino
flavor conversions in the cores of collapsing stars.

\section*{Acknowledgments}
We are grateful to Georg Raffelt for discussions during
the development of the project, to 
Shaoming Zhang for the 3D visualizations 
shown in Fig.~\ref{fig:3Drenderings}, and to Irene Tamborra for comments
on the manuscript. 
The work of F.C.\ is partially supported by the Deutsche
Forschungsgemeinschaft through Grants
SFB-1258 ``Neutrinos and Dark Matter in Astro- and Particle Physics
(NDM)'' and EXC~2094 ``ORIGINS: From the Origin
of the Universe to the First Building Blocks of Life'' 
as well as by the European Union
through Grant No.\ H2020-MSCA-ITN-2015/674896 (Innovative Training Network
``Elusives'').
M.S.\ acknowledges support from the National Science Foundation, 
Grant PHY-1630782, and from the Heising-Simons Foundation, Grant~2017-228.
The work of B.D.\ is partially supported by the Dept.\ of Atomic Energy
(Govt.\ of India) research project 12-R\&D-TFR-5.02-0200, the Dept.\ of
Science and Technology (Govt.\ of India) through a Ramanujan Fellowship,
and by the Max-Planck-Gesellschaft through a Max Planck Partner Group.
The work of A.M.\ is partially
supported by the Italian Istituto Nazionale di Fisica Nucleare 
(INFN) through the ``Theoretical Astroparticle Physics'' project and by 
research grant number 2017W4HA7S ``NAT-NET: Neutrino and Astroparticle 
Theory Network'' under the program PRIN 2017 funded by the Italian Ministero
dell'Universit\`a e della Ricerca (MUR).
At Garching, funding by the
European Research Council through Grant ERC-AdG No.~341157-COCO2CASA
and by the Deutsche Forschungsgemeinschaft (DFG, German Research Foundation)
through Sonderforschungsbereich (Collaborative Research Centre)
SFB-1258 ``Neutrinos and Dark Matter in Astro- and Particle Physics
(NDM)'' and under Germany's Excellence Strategy through
Excellence Cluster ORIGINS (EXC-2094)---390783311 is acknowledged.
The work of G.S.\ is partially supported by the DFG under
Germany's Excellence Strategy through Excellence Cluster 
``Quantum Universe'' (EXC-2121)---390833306.
Computer resources for this project have been
provided by the Leibniz Supercomputing Centre (LRZ) under
LRZ project ID pr62za and by the Max Planck Computing and Data
Facility (MPCDF) on the HPC system Hydra.

\bibliographystyle{JHEP}
\bibliography{moments}

\end{document}